\documentclass[useAMS,usenatbib]{mn2e}
\usepackage{graphicx} 
\usepackage{times}
\usepackage{epsfig}
\usepackage{rotating}
\usepackage{sidecap}
\usepackage{longtable}
\usepackage{lscape}
\include{definitions}
\title[The relativistic precession model and GRO J1655-40]{Precise mass and spin measurements for a stellar-mass black hole through X-ray timing: the case of GRO J1655-40}
\author[S. Motta et al.]{S.E. Motta$^{1}$, T.M. Belloni$^2$, L.Stella$^3$, T. Mu\~noz-Darias$^4$, R. Fender$^4$\\
$^{1}$ESAC, European Space Astronomy Centre, Villanueva de la Ca\~nada, E-28692 Madrid, Spain\\
$^{2}$INAF-Osservatorio Astronomico di Brera, Via E. Bianchi 46, I-23807 Merate, Italy\\
$^{3}$INAF, Osservatorio Astronomico di Roma, Via Frascati 33, I-00040, Monteporzio Catone, Italy\\
$^{4}$University of Southampton, Southampton, Hampshire, SO17 1BJ, United Kingdom\\
}

\begin{document}
\maketitle
\begin{abstract}

We present a systematic analysis of the fast time variability properties of the transient black hole binary GRO J1655-40, based on the complete set of \textit{Rossi XTE} observations. We demonstrate that the frequencies of the quasi periodic oscillations and of the broad band noise components and their variations match accurately the strong field general relativistic  frequencies of particle motion in the close vicinity of the innermost stable circular orbit, as predicted by the relativistic precession model.We obtain high precision measurements of the black hole mass (M = (5.31 $\pm$ 0.07) M$\odot$, consistent with the value from optical/NIR observations) and spin (a = 0.290 $\pm$ 0.003), through the sole use of X-ray timing.

\end{abstract}

\begin{keywords}
Black hole - accretion disks - binaries: close - stars: individual: GRO J1655-40 - X-rays: stars
\end{keywords}

%----------------------------------------------------------------------------------

\section{Introduction}

Quasi periodic oscillations (QPOs) in the flux emitted from accreting compact objects are a fairly common phenomenon and they are thought to originate in the innermost regions of the accretion flow. In a power density spectrum (PDS) they take the form of relatively narrow peaks yielding accurate centroid frequencies that can be associated with motion and/or accretion-related timescales in the strong gravitational field regime.
Despite  the fact that QPOs have been known for several decades, their origin is still not understood, and there is no consensus about their physical nature. However, several models have been proposed over the years (e.g., \citealt{Esin1997}, \citealt{Titarchuk1999}, \citealt{Tagger1999}, \citealt{Stella1998}, \citealt{Lamb2001},  \citealt{Abramowicz2001}, \citealt{Ingram2011} and references therein), several of them involving the predictions of the Theory of General Relativity (GR) and the fundamental frequencies of motion.

%QPOs are a common feature in accreting compact objects and come in many shapes and flavors.  
In black-hole systems, low-frequency ($\sim$ 0.1--30 Hz) QPOs (LFQPOs) of different kinds (dubbed type-A, -B and -C QPOs, see e.g., \citealt{Casella2005} and \citealt{Motta2012} for detailed discussions) and broader peaked noise components (at $\sim$ 1--100 Hz) with varying centroid frequencies have been detected (see e.g., \citealt{Belloni2011} for a review). QPOs with even higher frequencies (up to 450 Hz) were also observed (see, e.g., \citealt{Strohmayer2001}), but only a small number of detections are available (\citealt{Belloni2012}) and only in one case two simultaneous and therefore different high-frequency QPOs (called lower and upper HFQPOs) have been firmly detected (\citealt{Strohmayer2001}).
In weakly-magnetic accreting neutron stars (NSs), at least four distinct classes of QPOs have been identified so far: normal branch oscillation (NBOs, in the range 6--15 Hz, \citealt{Middleditch1986}), the horizontal branch oscillation (HBOs, in the $\sim$10--70 Hz frequency range, \citealt{vanderKlis1985}), and the lower and upper kHz QPOs ($\sim$ 100--1250 Hz) that typically occur in a pair (\citealt{vanderklis1996}). 

Bound orbits of matter in a gravitational field are characterised  by three different frequencies: the orbital  frequency and the vertical and radial epicyclic frequencies. %The difference between the general relativistic values of these  frequencies and classical  equivalents has so far been measured only in the Solar System and in radio pulsar binary systems hosting two degenerate stars. In these systems  weak field expansions of General Relativity (GR) suffice (e.g. \citealt{Psaltis2008}). 
GR predicts that the motion of matter at distances from a few to tens of gravitational radii  (r$_g$ = GM/c$^2$) from black holes (BHs) carries the signature of yet untested strong-field gravity effects which are among the fundamental consequences of Einstein's theory.  QPOs provide the most promising prospects to measure such characteristic frequencies in the electromagnetic radiation emitted by the plasma constituting the accretion flow.

So far, many models have been proposed to describe HF QPOs of black-hole LMXBs and several authors attempted to use X-ray timing to measure/constrain the mass and the spin of a compact object in a binary system. The relativistic precession model (RPM), was originally proposed by Stella \& Vietri (1998,1999) to explain the origin and the behaviour of the LFQPO and kHz QPOs in NS X-ray binaries and was later extended to BHs (\citealt{Stella1999a}). The RPM associates three types of QPOs observable in the PDS of accreting compact objects  to a combination of the fundamental frequencies of particle motion. The nodal precession frequency (or Lense-Thirring frequency) is associated with LFQPOs that show substantial changes in frequency (either a type-C QPO in BHs and an HBO in NSs), while the periastron precession frequency and the orbital frequency are associated with the lower and upper HFQPO, respectively (or to the lower and upper kHz QPO in the case of NSs). Even though the phenomenology of  QPOs in BHs systems is not as rich as that of NS binaries, BHs provide a ``cleaner'' environment to test the motion of matter in very strong gravitational fields. This is  because they do not possess  a solid surface nor a stably-anchored magnetic field. %Furthermore, from a more practical point of view, the determination of the spin in BH does not depend on the choice of the equation of state of the star (\citealt{Stella1998}). 
However, up to now, the application of the RPM to BH binaries was precluded by the absence of a clear detection of three simultaneous frequencies.

\cite{Boutloukos2006} tested the RPM on the data of the binary system Cir X-1 and found that the behaviour of HFQPOs in this source is in good agreement with the predictions of the RPM as well as of the Alfv\'en wave oscillation models (see \citealt{Zhang2004}). Assuming negligible frame dragging, these authors estimated the mass of the compact objects harbored by Cir X-1 (2.2 $\pm$ 0.3 M$\odot$). \cite{Boutloukos2006}  also found that the  predictions of the modified beat-frequency model proposed by \cite{Lamb2001} did not match the data of Cir X-1 and therefore needed further modification to be able to interpret the observational results.

\cite{Strohmayer2001} reported the first detection of two simultaneous HFQPOs in the light-curves of GRO J1655-40 (the same detections used in this work) and successfully interpreted them as the result of relativistic precession of matter and orbital motion of matter around a Kerr BH. \cite{Strohmayer2001} tentatively associated the peak observed at $\sim$17 Hz to the Lense-Thirring precession and, assuming a BH mass of 7M$\odot$, he obtained a spin between 0.4 and 0.6. The lack of a secure classification of the low frequency feature simultaneous to the HFQPOs as a type-C QPOs, prevented an unambiguous application of the RPM and self-consistent measurement of the mass and the spin of this object. 

\cite{Abramowicz2001} and \cite{Kluzniak2001} introduced the epicyclic resonant model, which was later   studied extensively by them as well as by other authors. This model is based on the assumption that non-linear 1:2 or 1:3 resonance between orbital and radial epicyclic motion could produce the HFQPOs observed in both BH and NS binaries and used it to constrain the spin of GRO J1655-40, obtaining a value of  0.2--0.67 for a mass in the range 5.5 to 7.9 solar masses. 

\cite{Bambi2012} studied from the theoretical point of view the predictions of the resonance model proposed by \cite{Kluzniak2001} and extended previous results to the case of non-Kerr space-times. They compared their findings with the measurements from the modelling of the soft X-ray continuum in the energy spectra and found that, for Kerr BHs, the two approaches do not provide consistent results.
\cite{Kato2012} revised the prediction of a resonantly-excited disk-oscillation model, still based on a resonance mechanism (the warp resonant model, see, e.g., \citealt{Kato2008}), and used it to infer a spin measurement of three different BH binaries, including GRO J1655-40, for which they obtained a spin between 0.9 and 0.99 for masses between 5.1 and 5.7 solar masses. The predictions of both these models quite strongly depend on which resonance mode is chosen to describe the HFQPOs observed. Most importantly, any measurement of the spin coming from these models must rely on the mass measured through alternative methods (i.e. dynamical studies, when available). This necessarily introduces large uncertainties. 

From spectroscopic studies, BH spins can be measured through two different methods: the modelling of the continuum X-ray spectrum of the accretion disk (e.g., \citealt{McClintock2011} and \citealt{McClintock2013}) and the fitting  of strong reflection features (especially the Fe K$\alpha$ line) observable in the spectra of accreting BH binaries (e.g., \citealt{Miller2007} and \citealt{Reynolds2013}). Both methods rely upon identifying the inner radius of the accretion disk with the innermost stable circular orbit. In addition, for the spectral continuum method, one must also know the mass of the black hole, the inclination of the accretion disk (generally assumed equal to the inclination of the binary system), and the distance to the binary (\citealt{Shafee2006}). These two methods have been for a long time the only ones available to measure BH spins and both have been widely used and tested to many different sources, sometimes simultaneously. 
For instance, recent results from the continuum fitting method on Cyg X-1 show that it harbours an almost maximally rotating BH (spin $>$ 0.95, \citealt{Gou2011}). This estimate is consistent with the results coming from the modelling of the reflection features (spin 0.97$^{+0.014}_{-0.02}$, \citealt{Fabian2012}). \citealt{Steiner2011} tested both methods on the BH binary XTE J1550-564 and obtained consistent results (spin  between -0.11 and 0.71 from the continuum method and spin 0.55$^{+0.15} _{- 0.22}$ from the iron line method). 
Estimates from both methods also exist for GRO J1655-40. \cite{Shafee2006}, modelling the thermal spectral continuum in both \textit{RXTE} and \textit{ASCA} data, found a spin of 0.65--0.75. \cite{Reis2009} fitted the strong reflection features in the GRO J1655-40 spectra in XMM-Newton data and found a lower limit for the spin of 0.90. 

GRO J1655-40 is among the few BH binary systems where an accurate measurement of the mass through dynamical studies based on optical and infrared observations has been obtained (\citealt{Beer2002}) and it is also the sole system where two simultaneous HFQPOs were clearly detected (\citealt{Strohmayer2001}).
In this work, we show that the predictions of the RPM match accurately the behaviour of the timing features observed in the BH binary GRO J1655-40 and, applying the RPM to the data of this source, we demonstrate that it is possible to measure with unprecedented precision the mass and spin of BHs in X-ray binaries through the sole use of X-ray timing.

%----------------------------------------------------------------------------------
\section{Observations and data analysis}\label{sec:observations} 

We examined a total of 571 Rossi X-ray Timing Explorer (\textit{RXTE})/Proportional Counter Array (\textit{PCA}) archival\footnote{http://heasarc.gsfc.nasa 5.gov/docs/xte/archive.html} observations of GRO J1655-40 obtained between 1996 March 14 (MJD 50157) and 2005 October 31 (MJD 53675) for a total exposure of $\sim$2.5Ms. Most of the data are concentrated between May 1996 and August 1997 and February 2005 and November 2005, during the two major outbursts of GRO J1655-40 (one in 1996, \citealt{Remillard1996} and one in 2005, \citealt{Markwardt2005}).
Most of these observations were already analysed in previous works (see, e.g., \citealt{Hjellming1995}, \citealt{Kuulkers1998}, \citealt{Greene2001}, \citealt{Beer2002}, \citealt{Remillard1996}, \citealt{Strohmayer2001}, \citealt{Remillard1999}, \citealt{Sobczak2000}, \citealt{Saito2006}, \citealt{Debnath2008} and others). %There are two main reasons why we re-analized all the available RXTE data of GRO J1655-40:
%\begin{itemize}
%\item the existing studies performed on this source lack of completeness and homogeneity, which are  needed in comprehensive studies such the one we present in this work. Since we are especially interested in the detection of HFQPO pairs, a systematic analysis is particularly important, as the only comprehensive timing study of GRO J1655-40 (\citealt{Belloni2012}) is based on a precedure aimed only at detecting one peak in each PDS.
%\item the characteristic frequencies of the features in the PDS must be measured with the highest possible precision, as the uncertainty on the frequency is the sole source of uncertainties on the final measurements of the BH parameters (see \ref{sec:RPM_solve}). In previous works this was not always a requirement.
%\end{itemize}

%The PCA data that we consider are now part of the public RXTE archive (http://heasarc.gsfc.nasa 5.gov/docs/xte/archive.html). 
The PCA data modes employed for most of these observations include a single-bit mode covering the lower energy events collected in the absolute channel range 0--35 and a high-time-resolution event mode recording events above the PCA absolute channel 36. We considered three energy bands for our analysis: 1.51-27.40 keV, 1.51-9.52 keV and 9.52-27.40 keV (corresponding to {\sc Standard 2} channels 1–-73, 1--31 and 32--73 at the beginning of the \textit{RXTE} mission). We will refer to these as the total, soft and hard bands, respectively. For some observations, these exact bands were not available due to the data modes, in which case we chose the closest approximation to those boundaries. Since the channel-energy relationship for \textit{RXTE/PCA} data changed over the years (see http://heasarc.nasa.gov/docs/xte/e-c\_table.html), four gain epochs have been defined. For this reason, for observations collected in different epochs we produced PDS in different channel-bands in order to ensure that all the PDS were produced in the same energy bands.

For each observation  we computed power spectra using custom software under \textsc{IDL}\footnote{http://www.brera.inaf.it/utenti/belloni/GHATS\_Package/Home.html} in the total, soft and hard energy band. We used 32s-long, 64s-long and 128s-long intervals of the event files and a Nyquist frequency of 2048 Hz to produce PDS. Then we averaged the individual spectra for each observation obtaining three different average PDS covering the frequency ranges 0.03-2048 Hz, 0.015-2048Hz and  0.008-2048 Hz and with frequency resolution $\sim$0.03 Hz, 0.015 Hz and 0.008 Hz, respectively. The PDS were normalised according to \cite{Leahy1983} and converted to square fractional rms (\citealt{Belloni1990}). We also measured the integrated fractional rms\footnote{We define the integrated fractional rms as the rms integrated over a certain frequency band.} in the 2-27 keV integrating the PDS over the 0.1--64 Hz frequency band and taking the square root of the power obtained. 
%The use of the Leahy normalization ensures that if the ligh-curve of a source consists only of Poisson fluctuations, then the periodogram is expected to be distributed exactly as $\chi^2$. This property makes the Leahy normalization the standard  for searching for periodic signals in the presence of Poisson noise and it is particularly suited for the search of high-frequency narrow features. This property also ensures the correctness of the fit performed in XSPEC through the $\chi^2$ statistics. 

To measure the hardness ratio, we used {\sc Standard 2} mode data, with a 16s-time resolution and suitable for the spectral analysis, to create background and dead-time corrected spectra.
We extracted energy spectra for each observation using the standard \textit{RXTE} software within \textsc{heasoft V. 6.12}. Only data coming from Proportional Counter Unit 2 (PCU2) of the PCA were used for the analysis, as it was the only unit that was active during all the observations. A systematic error of $0.6\%$ was added to the PCU2 spectra to account for calibration uncertainties\footnote{See http://www.universe.nasa.gov/xrays/programs/rxte/pca/doc/rmf/pcarmf-11.7 for a detailed discussion on the PCA calibration issues.}. 
We accumulated background corrected PCU2 rates in the {\sc Standard 2}\footnote{We refer to the energy bands valid in epoch 5.} channel bands A = 4 - 44 (3.3 - 20.2 keV), B = 4 - 10 ( 3.3 - 6.1 keV) and C = 11 - 20 (6.1 - 10.2 keV). The hardness is defined as H = C/B (\citealt{Homan2005a}, \citealt{Munoz-Darias2010}).

\subsection{PDS Fitting}\label{sec:PDS_fit}

The features we intend to analyse are found at different frequencies and must be treated differently. 
\begin{itemize}

\item Low frequency QPOs (LFQPOs) are usually found between $\sim$ 0.1 and 30Hz and are easily detected thanks to their intrinsically large rms amplitude. To fit them, we used PDS produced in the soft energy band from 32s-long intervals and we applied a logarithmical rebinning in such a way that each frequency bin was larger than the previous one by $\sim$2 per cent. 

\item Broad power-spectral components at low frequencies commonly show characteristic frequency between 3 and 20 Hz. Being broad, the power of these components is spread over a large range of frequencies, therefore we fitted them using PDS produced from 128s-long intervals in the soft energy band in order to extend the minimum observable frequency to lower values.

\item High Frequency QPOs (HFQPOs) are found around 300Hz and 450Hz in GRO J1655-40 and their intrinsic rms amplitude is smaller than in LFQPOs. For this reason their detection can be problematic. To study the 300 Hz and 450Hz HFQPOs we used PDS produced in the soft and hard band respectively and we applied to the average PDS from 32s-long intervals a linear rebinning to obtain a frequency resolution of either 8Hz or 16Hz (rebinning factor 128 and 256, respectively).

\end{itemize}

In all the cases, PDS fitting was carried out with the standard {\sc xspec v.12.7.1} fitting package by using a one-to-one energy-frequency conversion and a unit response. Following \cite{Belloni2002}, we fitted the Leahy-normalized PDS with a number of broad Lorentzian shapes. 

\begin{itemize}

\item A zero-centered low-frequency Lorentzian (defined L$_{b}$, see \citealt{Belloni2002} and Fig. 2) adequately describes the low-frequency end (the flat top part) of the band-limited noise visible in all power spectra. 

\item We fit one or more additional Lorentzians in the region around the low frequency QPO. The profile of the low frequency QPO can be more complex  than a single narrow Lorentzian. It is often observed as a relatively narrow peak (defined L$_{lf}$) accompanied by one or more harmonics and sometimes as a broader peak (called L$_{h}$) with a characteristic frequency slightly larger than the one of L$_{lf}$. %Since we selected for our analysis only observations where the LFQPO can be classified as a Type-C QPO (see \ref{sec:QPO_cla}), the L$_{lf}$ component in the PDS of our sample is always associated to a type-C QPOs.

\item Often, two additional components (defined L$_{l}$ and L$_{u}$, see \citealt{Belloni2002} and Fig. 2) have to be added to fit the high frequency end of the band-limited noise (see Tab. \ref{tab:sampleA}). They usually appear as broad components, even though sometimes they can take the form of narrow, highly-coherent high frequency QPOs (see Tab. \ref{tab:sampleB}). When observed in the form of narrow, coherent peaks, we fitted these components with narrow Lorentzians, otherwise broad Lorentzians are used.

\item We added to the fits a flat power-law to take into account the contribution of the Poissonian noise. Depending on the cases, the slope of the constant component was either fixed at zero or consistent with zero. 

\end{itemize}

Based on the results of the fitting we excluded from the subsequent analysis any feature that could not be detected significantly (significance $\leq$ 3$\sigma$, \citealt{Boutelier2010}).% The single trial signiﬁcances of QPOs are given as the ratio of the integral of the power of the Lorentzian used to fit the QPO divided by the negative 1-sigma error on the integral of the power. As shown by \cite{Boutelier2010}, this leads to and underestimation of the true significance of the QPOs. The best fit parameters from all the observations that we considered are listed in Tab. \ref{tab:sampleA} and \ref{tab:sampleB}.

\subsection{QPOs classification}\label{sec:QPO_cla}

The RPM was firstly proposed to explain the origin of LFQPOs in the PDS of NS binaries (\citealt{Stella1998}). The RPM interprets the peaks at tens of Hz in terms of the nodal precession of a narrow region close to the inner edge of the disk, dominated by the Lense-Thirring effect (\citealt{Lense1918}). Peaks in the correct frequency range (the HBOs) have been detected in several accreting NS binaries - both Atoll and Z sources (\citealt{Hasinger1989}). % However, only in Z sources these peaks have been firmly identified with a particular type of QPO, the HBOs \cite{Stella1998}.  
In BH binaries the HBOs find their equivalent in the type-C QPOs (\citealt{Casella2005}), which are observed in PDS of nearly all the known galactic accreting BH binaries. Hence, in this work we will only consider LFQPOs classified as type-C QPOs.

Besides type-C QPOs, BH binaries commonly show two other types of LFQPOs, dubbed as type-A and B QPOs (\citealt{Wijnands1999}, \citealt{Casella2004}), which are thought to be intrinsically different from type-C QPOs (\citealt{Motta2011a}) and hence unsuited to be used in the framework of the RPM.
For this reason the classification of the LFQPOs is crucial to the purpose of this work. Therefore, we performed an accurate classification following the method outlined in \cite{Motta2012}. 
In order to directly compare our result to those by these authors, we used the same extraction parameters that they used, i.e. we used PDS in the soft energy band from 64s-long interval and with a Nyquist frequency of 1024 Hz and we measured the integrated fractional rms over the frequency range 0.1-64Hz between 2 and 27 keV. 

\begin{figure}
\begin{center}
\includegraphics[width=8.0cm]{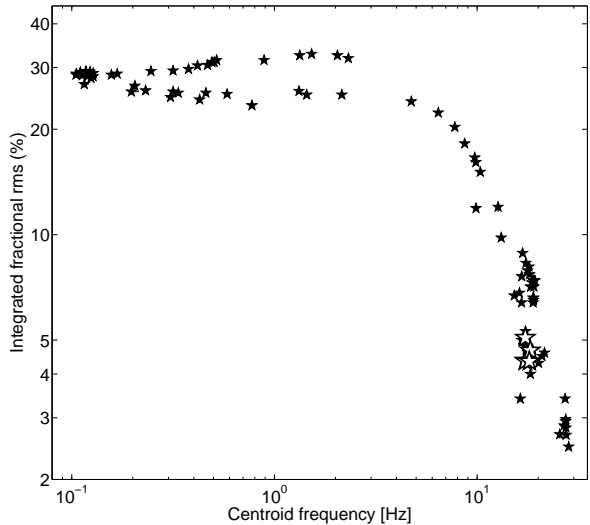}
\caption{QPO centroid frequency versus integrated (0.1–-64 Hz) fractional rms observed in GRO J1655-40 and plotted following Motta et al. (2012). All the observations of sample A and B are included and each point corresponds to a single \textit{RXTE} observation. The white stars correspond to the observations included in sample B, i.e. where a low frequency QPOs observed simultaneously with at least one HFQPO.}\label{fig:rms_freq}
\end{center}
\end{figure}

We fitted the PDS following the method outlined in Sec. \ref{sec:PDS_fit}. Following \cite{Casella2005} and \cite{Motta2011a,Motta2012}, we plot the integrated fractional rms of each PDS versus the centroid frequency of the QPO. This is a useful method for discriminating between different types of PDS as the integrated fractional rms is known to correlate well with the frequency of some type-C QPOs. In Fig. \ref{fig:rms_freq} we plot the integrated fractional rms as a function of the centroid frequency of the type-C QPO for all the observations of our sample. %Most of the points in Fig. \ref{fig:rms_freq} are already present in Fig. 5 of \cite{Motta2012}.

%All the PDS that we analyzed can be associated to PDS Type 2 described in \cite{Motta2012} and the LFQPO detected in there can  be solidly classified as a Type-C QPO. 

%----------------------------------------------------------------------------------
%----------------------------------------------------------------------------------

\subsection{Sample selection}\label{sec:selection}

We defined two different samples of observations that we analysed separately.

\begin{figure}
\begin{center}
\includegraphics[width=8.5cm]{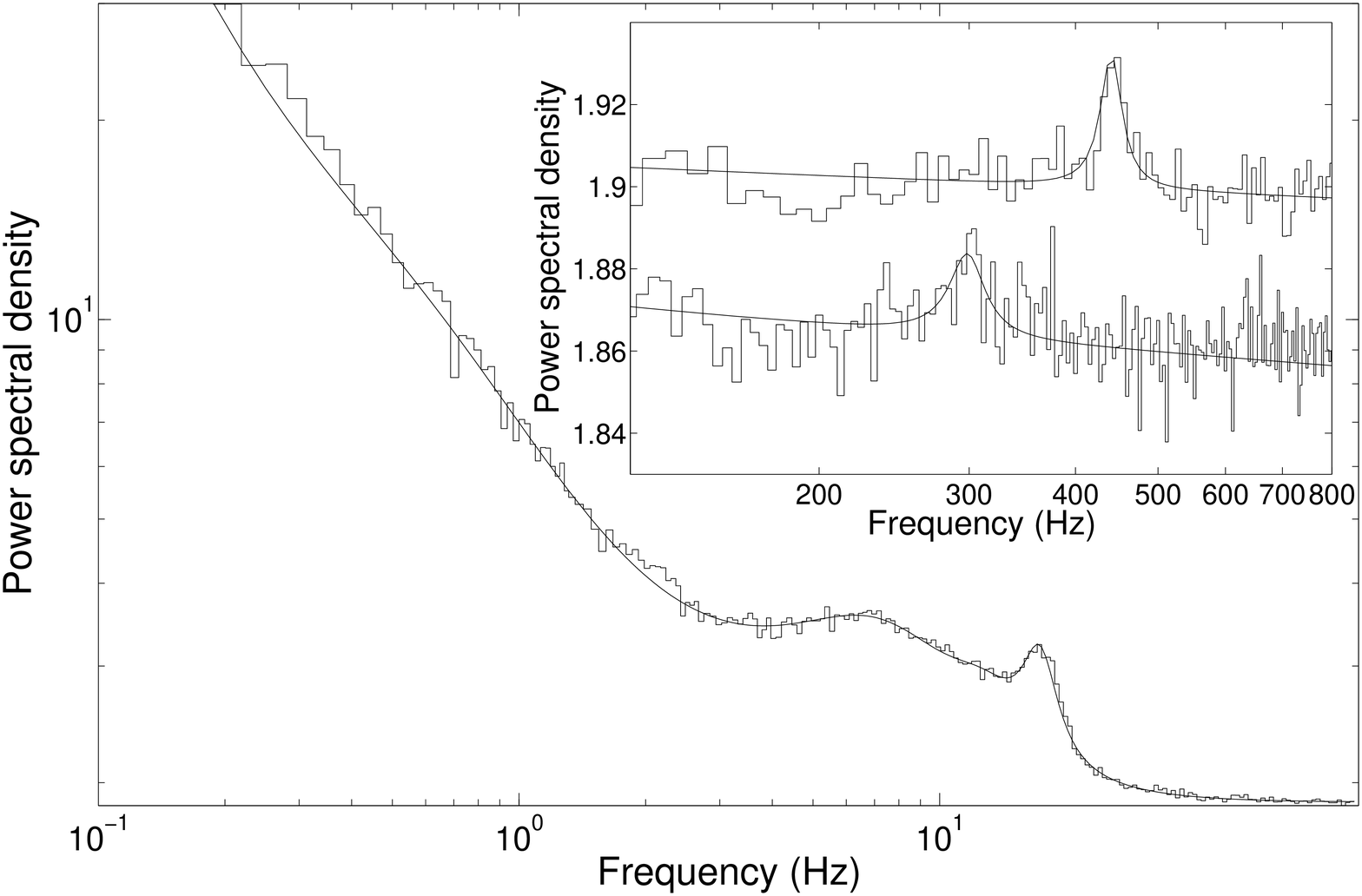}
\caption{PDS obtained averaging the observations of Sample B1. The figure shows the three simultaneous QPOs detected in the PDS. In the large panel we show the type-C QPO, while in the two insets we show the lower (top panel) and upper (bottom panel) HFQPOs.}\label{fig:PDS}
\end{center}
\end{figure}

\begin{itemize}

\item {\bf Sample A:} we considered only the observations whose PDS showed a narrow low-frequency feature that could be classified as a Type-C QPO. The final sample includes 94 type-C QPOs covering the $\sim$0.1-28 Hz frequency range, clearly detectable simultaneously in the soft, hard and total energy bands. Observations included in sample A are listed in  Tab. \ref{tab:sampleA}. Part of the QPOs included in this sample (all the QPOs coming for the outburst occurred in 2005) are also reported in \cite{Motta2012}, which covered only one of two outburst showed so far by GRO J1655-40.

\item {\bf Sample B:} we selected observations where we could detect at least a narrow feature at high frequencies (above 100 Hz) in the PDS. Then, we crossed this sample of observations with sample A in order to select only the observations showing both an HFQPO and a type-C LFQPO. In a few PDS we detected single HFQPOs, but either simultaneously to a different kind of LFQPO (a type-B) or no narrow low frequency feature. Those HFQPOs were excluded in this work.   
The final sample includes a total of 5 observations. 

\begin{itemize}

\item {\bf sub-sample B1:} three observations show a low frequency Type-C QPO at $\sim$17Hz, a HFQPO at $\sim$300 Hz and a HFQPO at $\sim$440 Hz simultaneously (sample B1); 

\item {\bf sub-sample B2:} two observations show a low frequency type-C QPO at $\sim$18 Hz and a HFQPO at $\sim$450 Hz. 

\end{itemize}

Lists of the HFQPOs detected in GRO J1655-40 are reported in \cite{Remillard1999}, \cite{Strohmayer2001} and \cite{Belloni2012}. Our sample B coincides with the sample obtained crossing the samples of these works.
In all the five observations residuals in the form of a QPO are visible at $\sim$300 Hz and/or $\sim$450 Hz in the PDS produced in the total energy band. However, the HFQPOs at $\sim$300 Hz are   detected in the soft band in three  observations, while the HFQPO at $\sim$450 Hz are detected in the hard energy band of the five observations. The type-C QPO is always clearly observable in the soft, hard and total  energy bands.
Since observations of sub-sample B1 show QPOs with consistent frequencies and since the source was in the same state (this can be inferred from the hardness ratio value and rms, see Tab. \ref{tab:sampleB}), in order to improve the quality of the PDS, following \cite{Strohmayer2001} we computed an average PDS by combining the three different observations. The resulting PDS is shown in Fig. \ref{fig:PDS}. We analysed separately the two observations of sub-sample B2 showing the Type-C QPO and the upper HFQPO. The properties of all the sample B observations are reported in Tab. \ref{tab:sampleB}. 

\end{itemize} 

%----------------------------------------------------------------------------------
%----------------------------------------------------------------------------------
%----------------------------------------------------------------------------------
\section{The relativistic precession model}\label{Sec:RPM}

%The RPM is based on basic predictions of General Relativity and was originally proposed by \cite{Stella1999}. 

%Astrophysical BHs can be described by an exact solution of General Relativity, discovered by \cite{Kerr1963}. This solution depends on two parameters: the mass and the angular momentum (or spin). Thanks to the complete integrability of geodesic motion in the Kerr metric, it is possible to calculate the orbits of a test particle in the gravitational field of the BH. 
%The Kerr metric expressed in the Boyder-Lindquist coordinates is given by (with G = c = 1):

%\begin{eqnarray}
%ds^2  &=&  -{{\Delta}\over {\Sigma}} \left(dt - a \sin^2\theta\ 
% d\phi\,\right)^2 + \Sigma \left(\frac{dr^2}{\Delta} + d\theta^{2}\right) +\nonumber \\
% %
% &+& \frac{\sin^2\theta}{\Sigma}
%\left[a dt - (r^2+a^2) \,d\phi \right]^2 \,
%\label{kerr}
%\end{eqnarray}
%where
%\begin{eqnarray}
%\Delta &=& r^2 + a^2 -2 M r\,,~~~~~~
%\Sigma= r^2+a^2 \cos^2\theta \nonumber\,
%\label{kmfunc}
%\end{eqnarray}
%$M$ is the mass of the BH and $a$ is the adimentional spin per unit mass of the BH $a = J/M^{2}$ (with $J$ angular momentum). The presence of the term $\Delta$ imposes the existence of an event horizon, whose radius is given by the largest root of the equation $\Delta$ = 0, 

%\begin{equation}
%r_{+}= M + \sqrt{M^2 - a^2 }\,
%\label{horizon+}
%\end{equation}
%and varies from M  to 2M over the range 0 $\leq$ a $\leq$ 1 and is defined only for a $ a \leq M$. $GM = r_{s}$ is the Schwartzchild radius of the BH.

In this work we apply the RPM (\citealt{Stella1998}, \citealt{Stella1999}), where certain combinations of the fundamental frequencies of motion in the strong field regime are associated with the frequency of certain QPOs observed in accreting compact objects. We adopt the convention G = c = 1.

When the motion occurs in the equatorial plane (\citealt{Bardeen1972}), from the geodesic equation we obtain the orbital frequency measured by a static observer at infinity:

\begin{equation}
\nu_{\phi} = \pm \frac{1}{2\pi} \left(\frac{M}{r^3}\right)^{1/2}\frac{1}{1 \pm a\left( \frac{M}{r}\right)^{3/2}}\label{nuphi}
\end{equation}

for a particle orbiting at a distance $r$ from a BH of mass $M$ and dimensionless spin parameter $a = J/M^{2}$ (with $J$ angular momentum and $J/M$ specific angular momentum).

Here $\pm \frac{1}{2\pi} \left(\frac{M}{r^3}\right)^{1/2}$ is the classical Keplerian frequency.
The upper sign always refers to the prograde orbits, while the lower sign refers to retrograde orbits. The off-equatorial (epicyclic) motion can be described applying a small perturbation in the circular (cyclic) orbit on the equatorial plane introducing velocity components in the r and $\theta$ directions (\citealt{Wilkins1972}). The resulting coordinate frequencies of the small amplitude radial oscillations within the plane (the epicyclic frequency $\nu_{r}$) and in the vertical direction (the vertical epicyclic frequency $\nu_{\theta}$ ) are given by:

\begin{equation}
\nu_{r} = \nu_{\phi} \left( 1-\frac{6 M}{r} - 3
a^2 \left(\frac{M}{ r}\right)^{2}  \pm \, 8 a \left(\frac{M}{ r}\right)^{3/2} \right)^{1/2}\label{nur}
\end{equation}
and 
\begin{equation}
\nu_{\theta} =  \nu_{\phi} \left(1
+3 a^2 \left(\frac{M}{ r}\right)^{2}  \mp \, 4 a \left(\frac{M}{r} 
\right)^{3/2} \right)^{1/2}\label{nuthe}
\end{equation}
These three coordinate frequencies lead to two additional frequencies, the periastron precession frequency: 

\begin{equation}
\nu_{per} = \nu_{\phi} - \nu_{r}\label{nupre}
\end{equation}
and the nodal precession frequency:

\begin{equation}
\nu_{nod} = \nu_{\phi} - \nu_{\theta}\label{nunod}
\end{equation}
The nodal precession frequency $\nu_{nod}$ is identically zero in the Swarzschild limit (a = 0), where the vertical epicyclic frequency $\nu_{\theta}$ equals $\nu_{\phi}$ (\citealt{Merloni1999}). The periastron precession frequency $\nu_{per}$ coincides with the orbital frequency $\nu_{\phi}$ at the radius of the innermost stable circular orbit, where the radial epicyclic frequency  equals zero. The innermost stable circular orbit is given by:

\begin{eqnarray}
r_{\rm ISCO} &=& M \left(3 + Z_2 \mp \left(\left(3- Z_1\right)\left(3 + Z_1 +2Z_2\right)\right)^{1/2} \right) \nonumber \\
Z_1 &=& 1 + \left(1-\frac{a^2}{r_g}\right)^{1/3}\left(\left(1+\frac{a}{r_g} \right)^{1/3} + \left(1-\frac{a}{r_g} \right)^{1/3} \right)  \nonumber \\
Z_2 &=& \left(\frac{3a^2}{r_g} + Z^2_1\right)^{1/2} 
\label{risco}
\end{eqnarray}
%where $r_{g}$ is the gravitational radius of the BH. 
Equations \ref{risco} are obtained requiring that the radial component of the gravitational potential and its derivative are identically zero (\citealt{Bardeen1972}). 

In the RPM the upper HFQPO is identified with the orbital frequency $\nu_{\phi}$ while the lower HFQPO is associated with the periastron precession frequency $\nu_{per}$. In the originally proposed version of the RPM (applied to the case of NSs, \citealt{Stella1999}) the LFQPO was associated with the second harmonic of the nodal precession frequency, $2 \nu_{nod}$, under the assumption that the inner accretion disk could be tilted in a way that a stronger signal could be produced at even harmonics of the nodal precession frequency  (\citealt{Psaltis1999}). Here, we use a simpler assumption and we associate the LFQPO frequency to the fundamental of the nodal precession frequency $\nu_{nod}$. 

Hence, under the assumption that the nodal precession frequency, the periastron precession frequency and the orbital frequency arise from the same radius, the system of equations that expresses the RPM is the following:

\begin{eqnarray}
\nu_{\phi} &=&  \pm \frac{1}{2\pi} \left(\frac{M}{r^3}\right)^{1/2}\frac{1}{1 \pm a\left( \frac{M}{r}\right)^{3/2}} \nonumber \\
\nu_{per} &=& v_{\phi}\left(1-\left( 1-\frac{6 M}{r} - 3
a^2 \left(\frac{M}{ r}\right)^{2}  \pm \, 8 a \left(\frac{M}{ r}\right)^{3/2} \right) ^{1/2}\right) \nonumber \\
\nu_{nod} &=& v_{\phi}\left(1-\left(1
+3 a^2 \left(\frac{M}{ r}\right)^{2}  \mp \, 4 a \left(\frac{M}{r} 
\right)^{3/2} \right) ^{1/2}\right)\label{RPMsys}
\end{eqnarray}

\section{Results}

\subsection{From the relativistic precession model to the spin and mass measurement}\label{sec:RPM_solve}

Equations \ref{RPMsys} show that the functional form of the frequencies expressing the RPM depends solely on the mass and the spin of the compact object and on the radius at which the QPOs are produced. Assuming that the three frequencies are produced at the same radius (\citealt{Stella1998}), if the three frequencies of the RPM are observed simultaneously, the system of equations of the RPM systems can be solved exactly. 
However, the RPM system is transcendental and cannot be solved analytically.   Therefore, we solved the RPM system numerically using the Newton method separately on the three equations following the steps outlined below. 

\begin{figure}
\begin{center}
%\begin{rotate}{90}
\includegraphics[width=8.5cm]{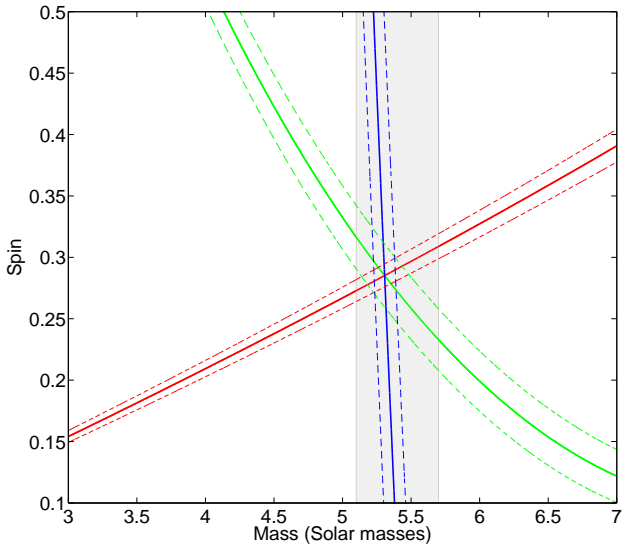}
\caption{BH spin as a function of the mass as predicted by the three equations of the relativistic precession model for the BH binary GRO J1655-40. The derived BH parameters are mass =  5.31 $\pm$ 0.07 M$_{\odot}$ and spin  a = 0.290 $\pm$ 0.003.
The green, red and blue lines represent the spin as a function of the mass according to the functional form of the nodal precession, the periastron precession and the orbital frequency, respectively. The solid lines mark the measured values and the dashed lines mark the 1-sigma confidence level on the mass measurement. The grey band marks the mass at 1-sigma confidence level, measured  independently from optical observations \citep{Beer2002}.}\label{fig:intersezione}
%\end{rotate}
\end{center}
\end{figure}

\begin{figure}
\begin{center}
\includegraphics[width=8.5cm]{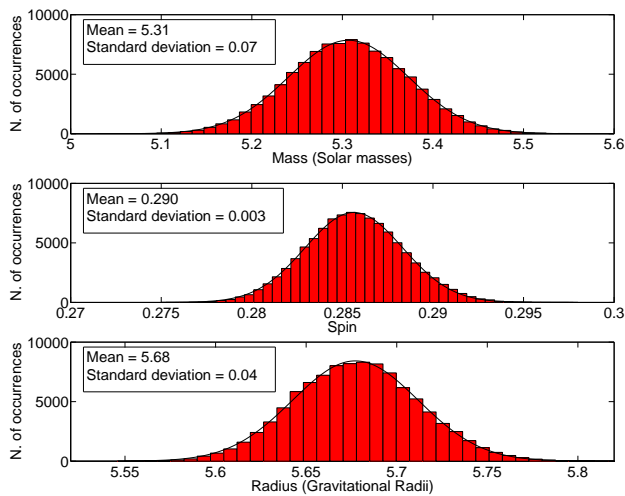}
\caption{From top to bottom, distributions of mass, spin and emission radius values obtained through the RPM and a Monte-Carlo simulation. }\label{fig:distributions}
\end{center}
\end{figure}

\begin{enumerate}

\item  For each of the RPM equations we use the Newton method to calculate the radius at which the corresponding observed QPO frequency (i.e. type-C QPO for the nodal precession frequency, lower HFQPO for the periastron precession frequency and upper HFQPO for the orbital frequency) is produced for every possible pair mass-spin in a given range. We considered masses between 3 and 50 solar masses with a resolution of 0.001 solar masses and spins between 0 and 1  with a resolution of 0.0005. This results in three independent sets of mass-spin-radius solutions, one for each RPM equation. 

\item We found the mass-spin-radius set of values that solves simultaneously the three equations. 

\item Through the Monte-Carlo method we simulated 10$^5$ sets of three frequencies based on the values measured from the PDS of GRO J1655-40 (see Sec. \ref{sec:PDS_fit} and Tab. \ref{tab:sampleB}). Each distribution is assumed to be Gaussian, centered at the centroid frequency of each QPOs and has a width equal to the error on the centroid frequency. 

\item We solved the RPM system for each set of three simulated frequencies following steps i and ii. We obtained three distributions of mass, spin and radius values (10$^5$ values for each distribution) consistent with being Gaussian-distributed.

\item Fitting the distribution of mass, spin and radius obtained in step iv (see fig. \ref{fig:distributions}) we obtain the following measurements: M = (5.31$\pm$0.07) M$_{\odot}$, a = 0.290$\pm$0.003, R = (5.68$\pm$0.04) R$_g$. The best fit parameters of the distribution fits are shown in Tab. \ref{tab:distribution_par}. Figure \ref{fig:intersezione} shows the graphical resolution of the RPM system for the averaged values of spin, mass and radius obtained through the Monte-Carlo simulation.  

\item From the spin and mass distribution we obtained a distribution of the measurements of r$_{\rm ISCO}$ and its corresponding nodal frequency according to Eq. \ref{RPMsys}  ($\nu_{nod}$). %These distribution are shown in Fig. \ref{fig:ISCO_dist} and 
The best fit parameters of the distribution fits are shown in Tab. \ref{tab:distribution_par}.  The mean value for r$_{\rm ISCO}$ is 5.031 $\pm$0.009 R$_g$.

\end{enumerate}

%----------------------------------------------------------------------------------
%----------------------------------------------------------------------------------
%----------------------------------------------------------------------------------
%----------------------------------------------------------------------------------
%----------------------------------------------------------------------------------
%\begin{figure}
%\begin{center}
%\includegraphics[width=8.5cm]{ISCO_distributions.eps}
%\caption{Top panel: Distributions of the R$_{ISCO}$ measures as obtained from the RPM and a Monte-Carlo %simulation. Bottom panel: corresponding distribution of the nodal precession frequency values at %R$_{ISCO}$. }\label{fig:ISCO_dist}
%\end{center}
%\end{figure}
%----------------------------------------------------------------------------------

%----------------------------------------------------------------------------------
%\subsection{Consistency checks of the RPM}

%----------------------------------------------------------------------------------
\subsection{The PBK correlation and the distribution of the type-C QPOs}\label{sec:PBK}

Psaltis, Belloni, van der Klis (1999), hereafter PBK, identified two components in the PDS of various BH and NS X-ray binaries, the frequencies of which follow a tight correlation over nearly three decades (the so-called PBK correlation). This correlation involves either two QPOs (a low-frequency QPO and either the lower or the upper HFQPO, dubbed as $L_l$ and $L_u$ in \citealt{Belloni2002}) or a low-frequency QPO (the component L$_{lf}$ in \citealt{Belloni2002}) and a broad noise component. 
\cite{Stella1999a} showed that the RPM can successfully explain the correlation found by \cite{Psaltis1999}. They showed that the dependence of the $L_l$  frequency on twice the $L_{lf}$ frequency matches the dependence of the periastron precession frequency on twice the nodal precession frequency.

Following \cite{Stella1999a}, we inspected the observations of GRO J1655-40 in  sample A to identify the power-spectral component following the PBK correlation (L$_{lf}$, L$_{l}$ and L$_{u}$ according to the nomenclature given above, based on \citealt{Belloni2002}). 

\begin{figure*}
\begin{center}
%\begin{rotate}{90}
\includegraphics[width=17.8cm]{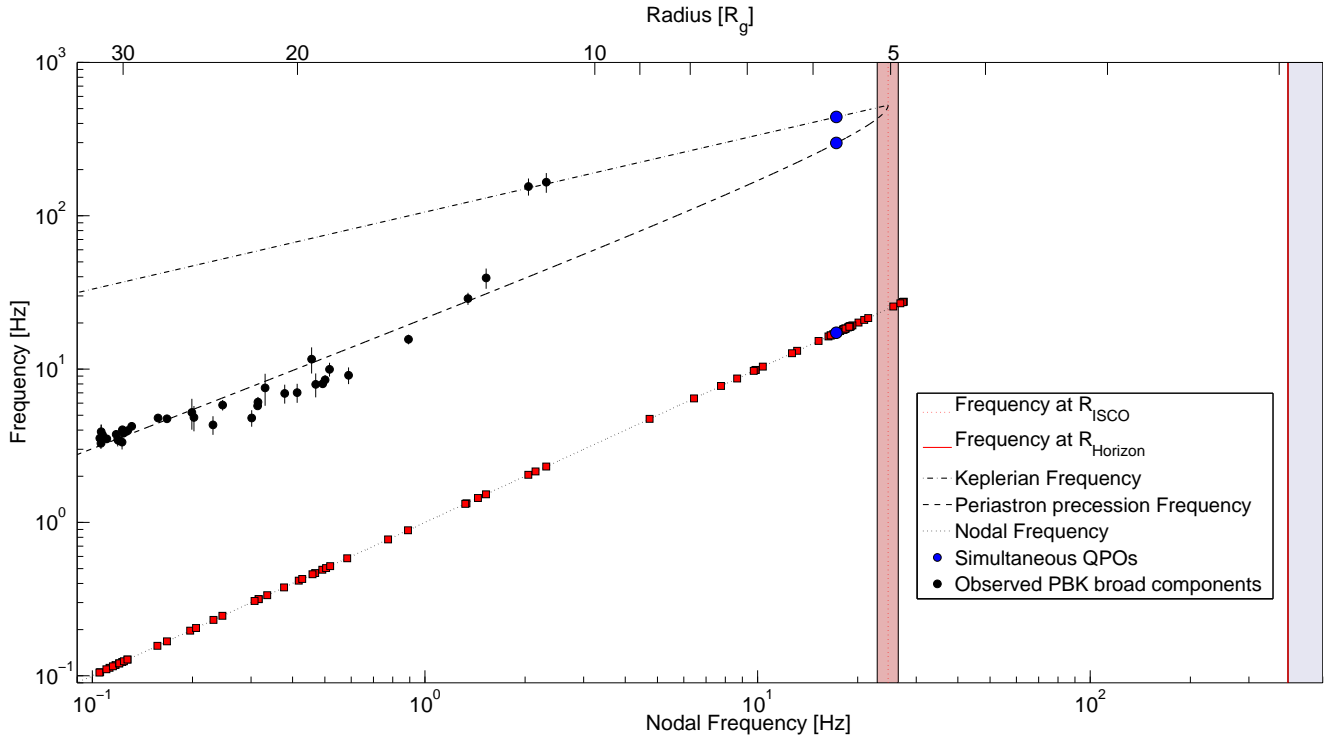}
\caption{Nodal precession frequency (dotted line), periastron precession frequency (dashed line) and  orbital frequency (dot-dashed line) as a function of the nodal precession frequency around a Kerr BH as predicted by the relativistic precession model. The lines are drawn for the mass and spin values (M = 5.31 M$\odot$ and a = 0.29) that provide the best fit to the three simultaneous QPO frequencies observed from GRO J1655-40 (blue points in the plot). The corresponding radii are given in the top X-axis. 
The black circles represent the characteristic frequencies of the broad components in the PDS of GRO 1655-40, which follow the PBK correlation. It is noteworthy that all points lie close to the low frequency extrapolation of the frequencies predicted by the RPM, based on the three simultaneous points only. 
The squares represent the frequency of type-C QPOs plotted against itself; this is to illustrate the frequency range over which these QPOs are detected, therefore their ``correlation'' is an artifact. All the points are plotted together with their 1-sigma error. When the error is not visible, it is smaller than the symbol.
The vertical dotted red line marks  the nodal frequency produced at the innermost stable circular orbit and the red vertical band indicates its corresponding 3-sigma uncertainty.}\label{fig:nu_vs_nu}
%\end{rotate}
\end{center}
\end{figure*}

We considered the characteristic frequency $\nu_{max}$ (defined as $\nu_{max}^2 = \nu^2 + (\Delta/2)^2$, where $\Delta$ is the width of the Lorentzian component describing a given power-spectral feature, see \citealt{Belloni2002}) of the components L$_{l}$ and L$_{u}$ and the peak frequency of the L$_{lf}$ component\footnote{The characteristic frequency $\nu_{max}$ constitute a measure for the break frequency of a broad Lorentzian and around this frequency the component contributes most of its power per logarithmic frequency interval. For the description of broad components $\nu_{max}$ is to be preferred to the Lorentzian peak frequency, since broad components are often centred at zero and the peaks frequency looses its meaning. Also, $\nu_{max}$ approaches the value of the peak frequency for decreasing widths of a Lorentzian component and it is practically coincident with the centroid frequency in the case of QPOs.}.
Following the prescriptions of the RPM (see Sec. \ref{Sec:RPM}), we plotted the characteristic frequencies L$_{l}$ and L$_{u}$ as a function of the L$_{lf}$ frequency. We also plotted the frequencies predicted by the RPM assuming the mass and spin obtained solving  the system as a function of the nodal frequency. The result is shown in Fig. \ref{fig:nu_vs_nu}.

\begin{itemize}

\item All the characteristic frequencies of the L$_{l}$ and L$_{u}$ components match well the frequencies predicted by the RPM. In particular, the dependence of the L$_{l}$ frequencies on the type-C QPO frequency (L$_{lf}$) follows the dependence of the periastron precession frequency on the nodal precession frequency, whereas the dependence of the L$_{u}$ frequencies on the type-C QPO frequencies matches the dependence of the orbital frequency on the nodal precession frequency (see Fig. \ref{fig:nu_vs_nu}). 

\item Most of the type-C QPOs in Sample A (L$_{lf}$ components, associated with the nodal precession motion) show characteristic frequencies that are consistent with being produced at radii larger than r$_{\rm ISCO}$. About 6\% of the detections are consistent with being produced at a radius which is slightly smaller than r$_{\rm ISCO}$, on an artificial extrapolation of the nodal precession frequency slightly inside r$_{\rm ISCO}$. The highest frequency Type-C QPO observed in the PDS of GRO J1655-40 is centered at 28.32 Hz would correspond to a radius equal to 4.8 gravitational radii ($\sim$4.5\% smaller than r$_{\rm ISCO}$). 

\end{itemize}

%----------------------------------------------------------------------------------
%----------------------------------------------------------------------------------
%----------------------------------------------------------------------------------
%\begin{figure}
%\begin{center}
%\includegraphics[width=8.5cm]{histo.eps}
%\caption{Distribution of type-C QPOs as observed in GRO J1655-40. The gray vertical band marks the 3-sigma level measurement of ISCO as obtained from the RPM, %(see Fig. \ref{fig:ISCO_dist}, bottom panel), 
%while the red vertical line corresponds to the Nodal precession frequency that would be produced at the event horizon.}\label{fig:Cdist}
%\end{center}
%\end{figure}
%---------------------------------------------------------------------------------

\subsection{On the width of the QPOs}\label{sec:width}

QPOs normally observed in BH X-ray binaries are typically narrow, with very small fractional widths ( $\Delta \nu$/$\nu \sim 10^{-1}$, \citealt{vanderKlis1997}). Such small fractional widths provide additional constraints on the physical mechanism that gives rise to the QPOs. The simplest assumption that we can make is that the QPOs are produced in a narrow annulus in the accretion flow. To obtain a rough estimate of the radial size $dr$ of this annulus, we apply to the emission radius at which the QPOs are produced a jitter as large as $dr$.
%Here we test the hypothesis that the width of the three simultaneous QPOs that we used to solve the RPM system is compatible with a jitter in the emission radius. Verifying this hypothesis supports the validity of the RPM and an suggests that the QPOs observed in accreting BH binaries could be the result of a modulation in frequency taking place at a timescale significantly shorter than a few seconds, i.e. the minimum cumulative exposure time that allows to produce a PDS with a good enough signal to noise ratio.
Using the mass, spin and emission radius values that we obtained from the RPM, we simulated the width of the three simultaneous QPOs that we used to solve the RPM system. We proceed as follows.

\begin{enumerate}

\item We allowed the emission radius to jitter within a certain, very small $dr$. We simulated a random jitter producing a normal distribution of 10$^5$ elements centred at zero with standard deviation equal to $dr$.  Then, we applied the jitter to the emission radius obtaining a distribution of emission radii.

\item We measured the nodal precession frequency, the periastron precession frequency and the orbital frequency using Eq. \ref{RPMsys} keeping mass and spin fixed at the values given in Sec. \ref{sec:RPM_solve} and varying the radius following the distribution described in step i. We obtain three distributions centred at the frequencies of the QPO we used to solve the RPM system.   

\item  The distributions that we obtain are slightly skewed and are well described by a log-normal distribution\footnote{We note here that the log-normal nature of the simulated distribution is probably an artifact due to the non-linear dependence of the RPM equations on the radius and on the assumption that the jitter is normally distributed around an average value. Given the intrinsic asymmetry of the space-time around a BH with respect to a certain radius, there is no specific reason to assume a symetric jitter. This choice is based on the lack of any better guess on how the jitter would look like to a static observer at infinity.}, while the real profile of the peaks in the PDS is Lorentzian or Normal. However, since the FWHM of a peaked distribution does not depend strongly on the shape of the distribution itself, we measured the FWHM of the distribution and we compared them with the FWHM of the three simultaneous observed QPOs. 

\item We increased $dr$ until we obtained distributions with FWHM consistent with the FWHM of the observed QPOs.

\end{enumerate}

%The distributions that we obtained are shown in Fig. \ref{fig:widths}. 
A jitter between 1.75\% and 2.4\% of the emission radius is able to reproduce the widths of the QPOs. 
The width of QPO associated with the nodal precession frequency (type-C QPO) lies between $\sim$2 and $\sim$3Hz, the width of the QPO associated with the periastron precession frequency (lower HFQPO) is found between  $\sim$42 and $\sim$58 Hz and the width of the QPO associated with the orbital frequency (upper HFQPO) is between $\sim$27 and $\sim$38 Hz. 
Table \ref{tab:confronto_width} summarises these results.
\section{Discussion}\label{sec:discussion}

The measurement of BH masses is a major issue in astrophysics. Even more problematic is the measure of their spin, which has been assumed to be null for a long time, owing to the lack of a measuring tool.  
Accurate mass measurements can be performed for a limited number of binary systems (about fifteen) with particular characteristics (i.e. detectable companion star and relatively high orbital inclination) through dynamical studies of their binary motion requiring long and often difficult multi-wavelength observations (see \citealt{Kreidberg2012}). 
Spins are obtained from spectroscopic studies through two different methods, both of which rely upon identifying the inner radius of the accretion disk with the innermost stable circular orbit, whose radius depends both on the mass and spin of
the black hole. The Fe K$\alpha$ line method is based on the modelling of the relativistically-broadened profile of the iron line (see \citealt{Miller2007} and \citealt{Reynolds2013} for recent reviews and e.g., \citealt{Miniutti2004}, \citealt{Brenneman2006}, \citealt{Yamada2009}, \citealt{Reis2011}, \citealt{Steiner2012}, \citealt{Miller2013}), while in the continuum-fitting method the thermal X-ray continuum spectrum of the accretion disk is modelled (\citealt{McClintock2011} and \citealt{McClintock2013} for recent reviews and e.g., \citealt{Shafee2006}, \citealt{Liu2008},\citealt{Gou2009}, \citealt{Steiner2010}, \citealt{Steiner2011}, \citealt{Steiner2012a}). %Both these methods often lead to measurements affected by relatively large uncertainties (see e.g. \citealt{Bambi2012}). 
 
The application of the RPM to X-ray high time-resolution data of BH binaries provides an independent  tool to measure at the same time the mass and the spin of a black hole, based solely on X-ray timing. We have shown that when three simultaneous QPOs\footnote{It is noteworthy that \textit{any} set of QPOs of the types relevant for the RPM, when observed simultaneously, would allow to solve the RPM system of equations.} (a low-frequency one and two high-frequency ones) are identified in the X-ray light-curves of a BH binary, their frequencies can be used to solve exactly the system of equations of the RPM, whose functional form depends solely on the mass and spin of the compact object, and the radius at which the orbiting matter gives rise to the QPOs. 
%This method rests on basic predictions of the theory of General Relativity (describing the orbital motion and the precession of matter around a Kerr BH), solely requires relatively simple and short X-ray observations and it is based only on the assumption that the three simultaneous QPOs originate from the same inhomogeneity in the accretion flow (see \citealt{Stella1998}). This assumption does not introduce large uncertainties (as, e.g. in the modelling of the spectral continuum and of the Iron line) and leads to solid and simultaneous measurements (not constraints) of both mass and spin. 

The three simultaneous oscillations required to apply this method to BH binaries have been observed only in the BH binary GRO J1655-40. These QPOs could be firmly identified on the basis of the results obtained by \cite{Motta2012} (who singled out the type-C QPO) and \cite{Strohmayer2001} (who firstly detected the two simultaneous HFQPOs).  
Applying the RPM to the QPOs from GRO J1655--40, we derived a mass equal to 5.31 $\pm$ 0.07 M$_{\odot}$, which is fully consistent with the value of the mass as determined independently through the most recent spectro-photometric optical observations (M$_{opt}$ = 5.4 $\pm$ 0.3 M$_{\odot}$, \citealt{Beer2002}, Fig. \ref{fig:intersezione}). The RPM also yields well-constrained values for the dimensionless spin per unit mass (a = 0.290 $\pm$ 0.003). Previous estimates of the spin of the BH in GRO 1655--40 were based on X-ray spectral analysis and gave higher, and yet different, values (e.g., a = 0.65 - 0.75, (\citealt{Shafee2006}); a = 0.94 - 0.98, (\citealt{Miller2009}) ; a = 0.9, (\citealt{Reis2009}) affected by larger uncertainties. 
The emission radius at the time when the three QPOs were observed was determined to be R = 5.68$\pm$ 0.03 Rg. Our method also succeeds in reproducing the width of each of the three simultaneous QPOs  by allowing for a $\sim$2\% jitter in the radius at which the QPOs are emitted (see Sec. \ref{sec:width}). This provides a rough estimate of the radial size of the region where the three simultaneous QPOs originated.   
%Identifying the exact reason for the radius to jitter is beyond the scope of this work. However, we argue that it could be due to an instability of the emission radius imposed by conditions unrelated to the mere effects of General Relativity, such as fluctuations in the accretion rate. 

The BH mass and the spin values obtained from the RPM allow us to predict the  expected behaviour of the frequencies  for each QPO type in GRO J1655--40  (see the lines in Fig. \ref{fig:nu_vs_nu}). All frequencies reach their highest allowed values at the innermost stable circular orbit, where the relativistic effective potential has an inflection point and the orbital and periastron precession frequencies coincide. Based on the mass and spin values that we measure, this is expected at a radius of 5.03 r$_g$ in GRO J1655--40, corresponding to a nodal precession frequency of 24.7 Hz. 
Once the RPM is solved for a set of three simultaneous frequencies, the identification of any of the three frequencies at other times allows us to measure the corresponding emission radius. In many of the \textit{RXTE} observations, GRO J1655--40 displayed low-frequency QPOs (Fig. \ref{fig:nu_vs_nu}) whose frequency varied over  a wide range, covering more than two decades, from $\sim$0.1 Hz to $\sim$28 Hz (\citealt{Motta2012}). In the nodal precession interpretation,  the lowest  frequency would be produced at $\sim$32 gravitational radii from the BH, while the highest would be arising from a radius about  4\% smaller than the inferred ISCO radius. This is in agreement with recent simulations that show that the accretion  disk might extend slightly inside the ISCO (\citealt{Krolik2002,Abramowicz2010,Penna2010}). %While Shakura-Sunyaev accretion disks have thickness equal to zero at $R_{ISCO}$ (Shakura \& Sunyaev 1973), real disks are thought to show a finite thickness everywhere. 
The minimum allowed radius for the inner edge of a real accretion disk is expected to deviate from the minimum allowed inner radius predicted for a Shakura-Sunyaev disk (corresponding to r$_{\rm ISCO}$) and to depend on the accretion rate and on the properties of the accretion flow (such as viscosity parameter, optical depth, \citealt{Abramowicz1988, Narayan1997, Krolik2002,Abramowicz2010,Penna2010}). 
In addition, it is worth noticing that what is meant by the \textit{inner edge} of an accretion disk around a BH depends on the property used to define the edge. Several different edges can be defined (\citealt{Krolik2002,Abramowicz2010,Penna2010}) and in all cases their minimum value lies in the vicinity of the ISCO. As noted by \cite{Krolik2002}, there is no reason why the minimum allowed values of any of these inner edges should coincide precisely with r$_{\rm ISCO}$, and according to the simulations, depending on which physical concept is under consideration, the minimum of any particular inner edge might be slightly inside or outside r$_{\rm ISCO}$. 
Moreover, being the location of the inner edge linked to the accretion rate, its position relative to r$_{\rm ISCO}$ is expected to change significantly with time.

The RPM also provides a natural interpretation for the PBK correlation (\citealt{Psaltis1999}). This correlation involves either two QPOs (a low-frequency QPO and either the lower or the upper HFQPO) or a low-frequency QPO and a broad noise component. For our case of study the periastron precession and orbital frequencies as a function of the nodal frequency match the correlation between the frequency of the type-C QPOs and the frequency of a broad noise components observed in GRO J1655--40 (see Fig. \ref{fig:nu_vs_nu}). This matching \textit{does not} involve any additional fitting to the data, but it qualitatively shows the agreement between the predictions of the model and the real behaviour of the timing features in GRO J1655-40. Moreover, it supports the hypothesis that both the QPO and broad noise components can be associated with the frequencies of the RPM (\citealt{Stella1999a}). A statistical analysis of the data aimed at demonstrating the consistency between the observed frequencies and the predicted ones is complex because of the transcendental nature of the RPM equations.
We are aware that several of the points have large scatter, which could be attributed to deviations of the behaviour of the matter in the accretion flow from the test-particle description (the investigation of the physical circumstances underlying the existence of these deviations will be presented in a forthcoming paper, see Motta et al. in prep.). 
However, we note that differently from QPOs (which are relatively narrow),  the precision with which we measure the characteristic frequencies of the broad PBK components is affected by the model used to fit them. This introduces additional uncertainties that could partly explain the large scatter of the points.

The application of the RPM that we have described allows us to make a significant step forward as it is currently the sole method able to provide self-consistent and simultaneous measurements of both mass \textit{and} spin without any \textit{a priori} knowledge or assumption of either of the two parameters.  
We note here that the simultaneous spin and mass measurements obtained through the RPM depend neither on the inclination of the source nor on its distance from the observer. Distance and inclination affect only the detectability of the QPOs and therefore the precision with which we can measure their frequencies.
However, we would like to stress that the RPM is a simplified model aimed at describing rather complex physical conditions, such as those encountered in the innermost disk regions around an accreting compact object. The RPM does not include yet a production mechanism for the QPO signals (but see \citealt{Psaltis2000} for a possible mechanism). This remains an open point to be investigated in the future. 

\section{Summary and conclusions}								
						
We reported the first successful attempt at obtaining precise simultaneous and self-consistent measurements of the mass and spin of a black hole in a binary system through the sole use of X-ray timing. 
Amongst BH binaries, GRO J1655-40 is the only source that has shown simultaneously the set of three quasi periodic oscillations that according to the relativistic precession model are to be associated with the frequency of motion predicted by the Theory of General Relativity. 
These frequencies allow us to solve exactly the system of equations described by the relativistic precession model and thus to obtain a simultaneous measurement of the mass and spin of the black hole, as well as the emission radius at which the oscillations are produced, with errors as small as $\sim$1\%. We obtained a mass (M = (5.31 $\pm$ 0.07) M$\odot$) which is fully consistent with the value obtained from optical/NIR dynamical studies. The spin that we obtain (a = 0.290 $\pm$ 0.003), however, is inconsistent with the estimates coming from either the Fe K$\alpha$ line method or the modelling of the spectral continuum method. 

The RPM can be successfully applied also to a few BH X-ray binaries showing at least two of the necessary set of three oscillations (Motta et al. in prep.). With future large area timing missions such as the LOFT satellite (which has been recently selected by ESA and it is now competing for a launch opportunity in 2020s) or the Indian mission ASTROSAT (currently planned for launch in 2014), the relativistic precession model might well become a very powerful tool to investigate many of the black hole binaries in our Galaxy.

\bigskip

\textit{
%\acknowledgments
%The authors would like to thank the anonymous referee for useful comments that contributed to improve the paper. 
SEM acknowledges the ESA research fellowship program and the University of Southampton for hospitality. SEM also acknowledges Dimitrios Emanuelopulos, Martin Henze, Erik Kuulkers, Anna Wolter, Sergio Campana and Andrea Merloni for useful suggestions and discussions. 
TMB and SEM acknowledge support from INAF PRIN 2012-6. TMD acknowledges funding via an EU Marie Curie Intra-European Fellowship under contract no. 2011-301355.
This research has made use of data obtained from the High Energy Astrophysics Science Archive Research Center (HEASARC), provided by NASA's Goddard Space Flight Center. 
%
%
%SEM acknowledges the little green alien throwing rocks into the accretion disc to produce QPOs. SEM also invites it to continue to do so for many years to come, possibly in a somehow predictable way as it seems it has been doing so far. He can choose the rules for himself as long as there is a chance for SEM to figure those out.
}

\newpage

\bibliographystyle{mn2e.bst}
\bibliography{biblio.bib}

\onecolumn																									
\begin{center}
\begin{landscape}
\begin{longtable}{|c c c c |c c |c c|} 																									
\caption{Observations included in Sample A (see Sec. \ref{sec:selection}). For each observation we report the Observation Id, the time at which the observation was taken, the integrated fractional rms (measured in the 2-27 keV energy band and in the 0.0-64.0 Hz frequency range), the hardness ration calculated as described in Sec. \ref{sec:observations}, the centroid frequency and the width of the Type-C QPO observed, the characteristic frequency ($\nu_{Max}$) of the broad high-frequency component - when detected - and the classification of such component. For details on the model used to fit the PDS, see Sec. \ref{sec:PDS_fit}}\label{tab:sampleA}. \\ 																									
\endfirsthead
\multicolumn{8}{c}{{\tablename\ \thetable{} -- continued from previous page}} \\ 
\hline								\multicolumn{4}{|c|}{ } & \multicolumn{2}{|c|}{Type-C QPOs} & \multicolumn{2}{|c|}{Broad components} \\				
\hline																\hline																							
Time	&	Obs. ID	&	Hardness ratio			&	rms 			&	$\nu$					&	$\Delta \nu$					&	$\nu_{Max}$					&	component type	\\
$[MJD]$	&		&				&	[\%]			&	[Hz]					&	[Hz]					&	[Hz]					&		\\

\hline																							\endhead																							
\hline 
\multicolumn{7}{c}{{Continued on next page}} \\											
\endfoot																							
\hline 																							\endlastfoot				
																							
\hline																									
\multicolumn{4}{|c|}{ } & \multicolumn{2}{|c|}{Type-C QPOs} & \multicolumn{2}{|c|}{Broad components} \\				\hline																					
\hline																									
Time	&	Obs. ID	&	Hardness ratio			&	rms 			&	$\nu$					&	$\Delta \nu$					&	$\nu_{Max}$					&	component type	\\
$[MJD]$	&		&				&	[\%]			&	[Hz]					&	[Hz]					&	[Hz]					&		\\																									
\hline
\hline																									
																									
50254.54	&	10261-01-05-00	&	0.459	$\pm$	0.002	&	8.9	$\pm$	0.18	&	16.8	$_{-	0.3	} ^{+	0.3	}$   &	5.7	$_{-	0.6	} ^{+	0.7	}$   &	-					&		\\
50254.61	&	10261-01-05-01	&	0.447	$\pm$	0.001	&	6.4	$\pm$	0.08	&	16.6	$_{-	0.2	} ^{+	0.2	}$   &	1.6	$_{-	0.6	} ^{+	0.8	}$   &	-					&		\\
50254.67	&	10261-01-05-02	&	0.445	$\pm$	0.001	&	6.8	$\pm$	0.07	&	16.2	$_{-	0.4	} ^{+	0.3	}$   &	3.4	$_{-	1.1	} ^{+	1.8	}$   &	-					&		\\
50289.36	&	10255-01-03-00 	&	0.412	$\pm$	0.001	&	4.0	$\pm$	0.1	&	18.3	$_{-	0.4	} ^{+	0.4	}$   &	4.9	$_{-	1.0	} ^{+	0.9	}$   &	-					&		\\
50310.60	&	10255-01-06-00 	&	0.404	$\pm$	0.001	&	4.3	$\pm$	0.1	&	20.1	$_{-	0.3	} ^{+	0.3	}$   &	3.9	$_{-	0.8	} ^{+	0.9	}$   &	-					&		\\
50311.39	&	10255-01-06-01	&	0.472	$\pm$	0.002	&	5.1	$\pm$	0.0	&	16.7	$_{-	0.1	} ^{+	0.1	}$   &	2.4	$_{-	0.3	} ^{+	0.3	}$   &	-					&		\\
50317.44	&	10255-01-07-00 	&	0.442	$\pm$	0.001	&	4.4	$\pm$	0.1	&	18.3	$_{-	0.1	} ^{+	0.1	}$   &	2.8	$_{-	0.3	} ^{+	0.3	}$   &	-					&		\\
50330.25	&	10255-01-09-00 	&	0.418	$\pm$	0.001	&	4.7	$\pm$	0.1	&	18.2	$_{-	0.1	} ^{+	0.1	}$   &	5.3	$_{-	0.4	} ^{+	0.4	}$   &	-					&		\\
50335.91	&	10255-01-10-00	&	0.428	$\pm$	0.001	&	4.4	$\pm$	0.1	&	18.1	$_{-	0.1	} ^{+	0.1	}$   &	4.3	$_{-	0.3	} ^{+	0.5	}$   &	-					&		\\
50346.20	&	10255-01-11-00 	&	0.397	$\pm$	0.001	&	4.5	$\pm$	0.1	&	20.9	$_{-	0.2	} ^{+	0.2	}$   &	4.8	$_{-	0.7	} ^{+	0.8	}$   &	-					&		\\
50378.08	&	10255-01-16-00 	&	0.415	$\pm$	0.001	&	4.6	$\pm$	0.1	&	21.5	$_{-	0.2	} ^{+	0.2	}$   &	5.8	$_{-	0.7	} ^{+	0.7	}$   &	-					&		\\
50383.56	&	10255-01-17-00 	&	0.461	$\pm$	0.001	&	4.4	$\pm$	0.1	&	17.4	$_{-	0.1	} ^{+	0.1	}$   &	4.1	$_{-	0.2	} ^{+	0.2	}$   &	-					&		\\
50389.22	&	10255-01-18-00 	&	0.575	$\pm$	0.002	&	11.9	$\pm$	0.1	&	9.87	$_{-	0.04	} ^{+	0.04	}$   &	2.2	$_{-	0.2	} ^{+	0.2	}$   &	-					&		\\
50394.88	&	20187-01-01-00 	&	0.514	$\pm$	0.002	&	3.4	$\pm$	0.1	&	16.3	$_{-	0.2	} ^{+	0.1	}$   &	2.5	$_{-	0.6	} ^{+	0.6	}$   &	-					&		\\
50512.76	&	20402-02-02-00	&	0.292	$\pm$	0.001	&	3.40	$\pm$	0.03	&	27.2	$_{-	0.2	} ^{+	0.1	}$   &	0.7	$_{-	0.5	} ^{+	0.5	}$   &	-					&		\\
50637.48	&	20402-02-20-00	&	0.273	$\pm$	0.001	&	2.48	$\pm$	0.04	&	28.3	$_{-	0.2	} ^{+	0.2	}$   &	1.5	$_{-	0.5	} ^{+	0.7	}$   &	-					&		\\
50674.44	&	20402-02-25-00 	&	0.702	$\pm$	0.003	&	22.3	$\pm$	0.1	&	6.44	$_{-	0.01	} ^{+	0.01	}$   &	0.62	$_{-	0.04	} ^{+	0.04	}$   &	-					&		\\
50678.57	&	20402-02-26-00 	&	0.863	$\pm$	0.006	&	23.4	$\pm$	0.3	&	0.77	$_{-	0.01	} ^{+	0.01	}$   &	0.20	$_{-	0.04	} ^{+	0.04	}$   &	-					&		\\
53427.02	&	90058-16-05-00	&	0.837	$\pm$	0.007	&	28.8	$\pm$	0.5	&	0.106	$_{-	0.002	} ^{+	0.002	}$   &	0.013	$_{-	0.003	} ^{+	0.006	}$   &	3.3	$_{-	0.3	} ^{+	0.3	}$ &	$L_{l}$	\\
53427.15	&	90428-01-01-01	&	0.825	$\pm$	0.006	&	28.6	$\pm$	0.4	&	0.105	$_{-	0.002	} ^{+	0.004	}$   &	0.029	$_{-	0.003	} ^{+	0.010	}$   &	3.7	$_{-	0.3	} ^{+	0.3	}$ &	$L_{l}$	\\
53427.94	&	90058-16-07-00	&	0.837	$\pm$	0.006	&	29.3	$\pm$	0.3	&	0.117	$_{-	0.001	} ^{+	0.004	}$   &	0.019	$_{-	0.003	} ^{+	0.005	}$   &	3.9	$_{-	0.4	} ^{+	0.4	}$ &	$L_{l}$	\\
53428.14	&	90428-01-01-03	&	0.831	$\pm$	0.006	&	28.5	$\pm$	0.4	&	0.113	$_{-	0.002	} ^{+	0.003	}$   &	0.025	$_{-	0.005	} ^{+	0.006	}$   &	3.9	$_{-	0.4	} ^{+	0.4	}$ &	$L_{l}$	\\
53428.86	&	90428-01-01-04	&	0.862	$\pm$	0.004	&	29.2	$\pm$	0.2	&	0.123	$_{-	0.001	} ^{+	0.001	}$   &	0.015	$_{-	0.001	} ^{+	0.005	}$   &	4.0	$_{-	0.2	} ^{+	0.2	}$ &	$L_{l}$	\\
53429.71	&	90428-01-01-02	&	0.837	$\pm$	0.004	&	28.9	$\pm$	0.2	&	0.128	$_{-	0.001	} ^{+	0.002	}$   &	0.018	$_{-	0.001	} ^{+	0.006	}$   &	4.0	$_{-	0.2	} ^{+	0.2	}$ &	$L_{l}$	\\
53430.96	&	90428-01-01-05	&	0.832	$\pm$	0.005	&	29.1	$\pm$	0.3	&	0.110	$_{-	0.002	} ^{+	0.003	}$   &	0.011	$_{-	0.003	} ^{+	0.006	}$   &	3.5	$_{-	0.2	} ^{+	0.2	}$ &	$L_{l}$	\\
53431.17	&	90058-16-06-00	&	0.834	$\pm$	0.008	&	26.9	$\pm$	0.6	&	0.115	$_{-	0.002	} ^{+	0.006	}$   &	0.024	$_{-	0.007	} ^{+	0.012	}$   &	3.7	$_{-	0.5	} ^{+	0.5	}$ &	$L_{l}$	\\
53431.61	&	90428-01-01-06	&	0.830	$\pm$	0.008	&	28.4	$\pm$	0.6	&	0.120	$_{-	0.005	} ^{+	0.002	}$   &	0.026	$_{-	0.004	} ^{+	0.008	}$   &	3.5	$_{-	0.3	} ^{+	0.4	}$ &	$L_{l}$	\\
53431.74	&	90428-01-01-07	&	0.869	$\pm$	0.007	&	28.6	$\pm$	0.4	&	0.121	$_{-	0.006	} ^{+	0.002	}$   &	0.032	$_{-	0.007	} ^{+	0.017	}$   &	3.4	$_{-	0.2	} ^{+	0.2	}$ &	$L_{l}$	\\
53431.81	&	90428-01-01-08	&	0.870	$\pm$	0.008	&	28.0	$\pm$	0.6	&	0.124	$_{-	0.005	} ^{+	0.002	}$   &	0.024	$_{-	0.006	} ^{+	0.013	}$   &	3.3	$_{-	0.3	} ^{+	0.5	}$ &	$L_{l}$	\\
53431.88	&	90428-01-01-09	&	0.853	$\pm$	0.004	&	28.3	$\pm$	0.2	&	0.128	$_{-	0.002	} ^{+	0.002	}$   &	0.034	$_{-	0.006	} ^{+	0.006	}$   &	4.2	$_{-	0.2	} ^{+	0.2	}$ &	$L_{l}$	\\
53432.79	&	90428-01-01-10	&	0.834	$\pm$	0.004	&	28.6	$\pm$	0.2	&	0.157	$_{-	0.001	} ^{+	0.002	}$   &	0.031	$_{-	0.002	} ^{+	0.006	}$   &	4.8	$_{-	0.3	} ^{+	0.3	}$ &	$L_{l}$	\\
53433.00	&	91404-01-01-00	&	0.833	$\pm$	0.005	&	28.8	$\pm$	0.3	&	0.168	$_{-	0.002	} ^{+	0.003	}$   &	0.025	$_{-	0.003	} ^{+	0.007	}$   &	4.7	$_{-	0.3	} ^{+	0.3	}$ &	$L_{l}$	\\
53433.91	&	91404-01-01-02	&	0.823	$\pm$	0.003	&	29.3	$\pm$	0.2	&	0.246	$_{-	0.002	} ^{+	0.002	}$   &	0.044	$_{-	0.002	} ^{+	0.009	}$   &	5.8	$_{-	0.5	} ^{+	0.9	}$ &	$L_{l}$	\\
53434.69	&	91404-01-01-03	&	0.857	$\pm$	0.005	&	29.4	$\pm$	0.3	&	0.316	$_{-	0.006	} ^{+	0.003	}$   &	0.08	$_{-	0.01	} ^{+	0.01	}$   &	5.8	$_{-	0.1	} ^{+	1.2	}$ &	$L_{l}$	\\
53435.61	&	91404-01-01-01	&	0.807	$\pm$	0.005	&	29.7	$\pm$	0.3	&	0.377	$_{-	0.004	} ^{+	0.004	}$   &	0.06	$_{-	0.01	} ^{+	0.01	}$   &	7	$_{-	1	} ^{+	1	}$ &	$L_{l}$	\\
53436.16	&	91404-01-01-04	&	0.806	$\pm$	0.005	&	30.4	$\pm$	0.3	&	0.417	$_{-	0.005	} ^{+	0.005	}$   &	0.07	$_{-	0.01	} ^{+	0.01	}$   &	7	$_{-	1	} ^{+	2	}$ &	$L_{l}$	\\
53436.40	&	91404-01-01-05	&	0.806	$\pm$	0.006	&	30.5	$\pm$	0.4	&	0.468	$_{-	0.006	} ^{+	0.010	}$   &	0.09	$_{-	0.02	} ^{+	0.02	}$   &	8	$_{-	1	} ^{+	3	}$ &	$L_{l}$	\\
53436.73	&	91702-01-01-00	&	0.801	$\pm$	0.003	&	31.1	$\pm$	0.1	&	0.491	$_{-	0.003	} ^{+	0.003	}$   &	0.083	$_{-	0.006	} ^{+	0.007	}$   &	8.0	$_{-	0.5	} ^{+	0.4	}$ &	$L_{l}$	\\
53437.07	&	91702-01-01-01	&	0.811	$\pm$	0.004	&	31.1	$\pm$	0.2	&	0.504	$_{-	0.005	} ^{+	0.005	}$   &	0.10	$_{-	0.01	} ^{+	0.01	}$   &	8.5	$_{-	0.6	} ^{+	0.7	}$ &	$L_{l}$	\\
53437.14	&	91702-01-01-02	&	0.806	$\pm$	0.004	&	31.5	$\pm$	0.3	&	0.519	$_{-	0.005	} ^{+	0.005	}$   &	0.09	$_{-	0.01	} ^{+	0.01	}$   &	10	$_{-	1	} ^{+	1	}$ &	$L_{l}$	\\
53438.06	&	91702-01-01-03	&	0.766	$\pm$	0.003	&	31.5	$\pm$	0.2	&	0.889	$_{-	0.005	} ^{+	0.005	}$   &	0.16	$_{-	0.01	} ^{+	0.01	}$   &	16	$_{-	1	} ^{+	1	}$ &	$L_{l}$	\\
53438.76	&	91702-01-01-04	&	0.749	$\pm$	0.003	&	32.5	$\pm$	0.2	&	1.333	$_{-	0.005	} ^{+	0.005	}$   &	0.17	$_{-	0.02	} ^{+	0.02	}$   &	29	$_{-	3	} ^{+	3	}$ &	$L_{l}$	\\
53439.11	&	91702-01-01-05	&	0.744	$\pm$	0.003	&	32.8	$\pm$	0.3	&	1.526	$_{-	0.005	} ^{+	0.005	}$   &	0.23	$_{-	0.02	} ^{+	0.02	}$   &	39	$_{-	6	} ^{+	7	}$ &	$L_{l}$	\\
53439.61	&	90704-04-01-01	&	0.718	$\pm$	0.003	&	32.5	$\pm$	0.2	&	2.045	$_{-	0.004	} ^{+	0.004	}$   &	0.27	$_{-	0.01	} ^{+	0.01	}$   &	155	$_{-	20	} ^{+	22	}$ &	$L_{u}$	\\
53439.74	&	90704-04-01-00	&	0.704	$\pm$	0.002	&	31.9	$\pm$	0.2	&	2.315	$_{-	0.004	} ^{+	0.004	}$   &	0.29	$_{-	0.01	} ^{+	0.01	}$   &	166	$_{-	24	} ^{+	31	}$ &	$L_{u}$	\\
53500.79	&	91702-01-52-03 	&	0.387	$\pm$	0.001	&	7.1	$\pm$	0.1	&	18.4	$_{-	0.3	} ^{+	0.3	}$   &	4.0	$_{-	0.6	} ^{+	0.7	}$   &	-					&		\\
53502.35	&	91702-01-54-00 	&	0.391	$\pm$	0.001	&	7.4	$\pm$	0.0	&	18.89	$_{-	0.07	} ^{+	0.07	}$   &	2.9	$_{-	0.2	} ^{+	0.2	}$   &	-					&		\\
53502.42	&	91702-01-54-01 	&	0.392	$\pm$	0.001	&	7.1	$\pm$	0.0	&	19.1	$_{-	0.1	} ^{+	0.1	}$   &	3.3	$_{-	0.3	} ^{+	0.4	}$   &	-					&		\\
53502.49	&	91702-01-54-02 	&	0.395	$\pm$	0.001	&	7.4	$\pm$	0.0	&	19.27	$_{-	0.09	} ^{+	0.09	}$   &	2.7	$_{-	0.3	} ^{+	0.3	}$   &	-					&		\\
53502.56	&	91702-01-54-03 	&	0.365	$\pm$	0.001	&	7.9	$\pm$	0.1	&	17.7	$_{-	0.3	} ^{+	0.3	}$   &	2.8	$_{-	1.0	} ^{+	1.0	}$   &	-					&		\\
53506.28	&	91702-01-56-00G 	&	0.416	$\pm$	0.001	&	8.1	$\pm$	0.0	&	18.04	$_{-	0.02	} ^{+	0.02	}$   &	2.35	$_{-	0.07	} ^{+	0.07	}$   &	-					&		\\
53506.95	&	91702-01-58-03 	&	0.397	$\pm$	0.001	&	6.6	$\pm$	0.1	&	19.0	$_{-	0.2	} ^{+	0.2	}$   &	4.4	$_{-	0.5	} ^{+	0.5	}$   &	-					&		\\
53507.02	&	91702-01-58-04 	&	0.397	$\pm$	0.001	&	7.3	$\pm$	0.1	&	18.7	$_{-	0.1	} ^{+	0.1	}$   &	3.3	$_{-	0.3	} ^{+	0.3	}$   &	-					&		\\
53507.09	&	91702-01-58-02 	&	0.382	$\pm$	0.001	&	6.4	$\pm$	0.1	&	18.9	$_{-	0.2	} ^{+	0.2	}$   &	3.7	$_{-	0.5	} ^{+	0.6	}$   &	-					&		\\
53507.20	&	91702-01-57-00G 	&	0.406	$\pm$	0.001	&	7.6	$\pm$	0.0	&	16.59	$_{-	0.02	} ^{+	0.02	}$   &	1.29	$_{-	0.06	} ^{+	0.06	}$   &	-					&		\\
53507.73	&	91702-01-59-00 	&	0.419	$\pm$	0.001	&	7.7	$\pm$	0.1	&	18.14	$_{-	0.06	} ^{+	0.06	}$   &	2.4	$_{-	0.2	} ^{+	0.2	}$   &	-					&		\\
53508.51	&	91702-01-58-00 	&	0.533	$\pm$	0.002	&	4.7	$\pm$	0.0	&	18.8	$_{-	0.2	} ^{+	0.2	}$   &	7.6	$_{-	0.5	} ^{+	0.5	}$   &	-					&		\\
53509.23	&	91702-01-59-02 	&	0.428	$\pm$	0.001	&	8.3	$\pm$	0.0	&	17.42	$_{-	0.02	} ^{+	0.02	}$   &	1.7	$_{-	0.0	} ^{+	0.1	}$   &	-					&		\\
53509.56	&	91702-01-58-01 	&	0.400	$\pm$	0.001	&	7.1	$\pm$	0.0	&	18.38	$_{-	0.05	} ^{+	0.05	}$   &	2.6	$_{-	0.1	} ^{+	0.1	}$   &	-					&		\\
53510.03	&	91702-01-60-02 	&	0.390	$\pm$	0.001	&	6.5	$\pm$	0.1	&	19.0	$_{-	0.2	} ^{+	0.2	}$   &	3.8	$_{-	0.5	} ^{+	0.5	}$   &	-					&		\\
53510.10	&	91702-01-60-00	&	0.361	$\pm$	0.001	&	6.7	$\pm$	0.0	&	15.3	$_{-	0.1	} ^{+	0.1	}$   &	8.0	$_{-	0.2	} ^{+	0.2	}$   &	-					&		\\
53510.28	&	91702-01-60-01 	&	0.374	$\pm$	0.001	&	6.6	$\pm$	0.1	&	18.9	$_{-	0.3	} ^{+	0.3	}$   &	4.1	$_{-	0.6	} ^{+	0.7	}$   &	-					&		\\
53515.20	&	91702-01-63-01 	&	0.370	$\pm$	0.001	&	5.1	$\pm$	0.0	&	16.9	$_{-	0.1	} ^{+	0.1	}$   &	5.1	$_{-	0.3	} ^{+	0.3	}$   &	-					&		\\
53515.66	&	91702-01-63-00 	&	0.366	$\pm$	0.001	&	5.3	$\pm$	0.0	&	17.3	$_{-	0.1	} ^{+	0.1	}$   &	5.0	$_{-	0.4	} ^{+	0.4	}$   &	-					&		\\
53516.25	&	91702-01-64-02 	&	0.353	$\pm$	0.001	&	4.4	$\pm$	0.0	&	17.3	$_{-	0.2	} ^{+	0.2	}$   &	3.4	$_{-	0.6	} ^{+	0.6	}$   &	-					&		\\
53516.45	&	91702-01-64-03 	&	0.351	$\pm$	0.001	&	4.8	$\pm$	0.0	&	17.5	$_{-	0.2	} ^{+	0.2	}$   &	3.2	$_{-	0.7	} ^{+	0.7	}$   &	-					&		\\
53516.57	&	91702-01-64-00 	&	0.369	$\pm$	0.001	&	4.8	$\pm$	0.0	&	17.29	$_{-	0.09	} ^{+	0.1	}$   &	4.5	$_{-	0.3	} ^{+	0.3	}$   &	-					&		\\
53517.04	&	91702-01-65-00 	&	0.372	$\pm$	0.001	&	5.0	$\pm$	0.0	&	16.6	$_{-	0.2	} ^{+	0.2	}$   &	5.4	$_{-	0.5	} ^{+	0.5	}$   &	-					&		\\
53517.10	&	91702-01-65-03 	&	0.347	$\pm$	0.001	&	4.7	$\pm$	0.1	&	17.5	$_{-	0.3	} ^{+	0.3	}$   &	4.6	$_{-	1.0	} ^{+	1.1	}$   &	-					&		\\
53574.40	&	91702-01-16-10	&	0.162	$\pm$	0.001	&	3.0	$\pm$	0.0	&	27.4	$_{-	0.2	} ^{+	0.2	}$   &	1.8	$_{-	0.5	} ^{+	0.6	}$   &	-					&		\\
53575.60	&	91702-01-17-10	&	0.172	$\pm$	0.001	&	2.7	$\pm$	0.1	&	27.5	$_{-	0.1	} ^{+	0.1	}$   &	0.7	$_{-	0.2	} ^{+	0.3	}$   &	-					&		\\
53580.40	&	91702-01-21-10	&	0.161	$\pm$	0.001	&	2.8	$\pm$	0.1	&	27.4	$_{-	0.4	} ^{+	0.4	}$   &	3.1	$_{-	0.7	} ^{+	0.9	}$   &	-					&		\\
53583.40	&	91702-01-24-10	&	0.158	$\pm$	0.001	&	2.9	$\pm$	0.1	&	27.3	$_{-	0.1	} ^{+	0.1	}$   &	1.0	$_{-	0.4	} ^{+	0.4	}$   &	-					&		\\
53585.40	&	91702-01-25-11	&	0.157	$\pm$	0.001	&	2.9	$\pm$	0.1	&	26.9	$_{-	0.2	} ^{+	0.2	}$   &	1.7	$_{-	0.4	} ^{+	0.5	}$   &	-					&		\\
53593.30	&	91702-01-32-10	&	0.147	$\pm$	0.001	&	2.7	$\pm$	0.1	&	25.6	$_{-	0.2	} ^{+	0.44	}$   &	1.4	$_{-	0.4	} ^{+	0.9	}$   &	-					&		\\
53628.20	&	91702-01-76-00	&	0.362	$\pm$	0.001	&	9.8	$\pm$	0.1	&	13.15	$_{-	0.05	} ^{+	0.05	}$   &	1.5	$_{-	0.1	} ^{+	0.2	}$   &	-					&		\\
53628.59	&	91702-01-76-01	&	0.407	$\pm$	0.002	&	12.0	$\pm$	0.4	&	12.69	$_{-	0.06	} ^{+	0.07	}$   &	0.8	$_{-	0.2	} ^{+	0.2	}$   &	-					&		\\
53628.92	&	91702-01-71-03	&	0.500	$\pm$	0.003	&	16.1	$\pm$	0.3	&	9.86	$_{-	0.01	} ^{+	0.01	}$   &	0.39	$_{-	0.05	} ^{+	0.05	}$   &	-					&		\\
53628.98	&	91702-01-71-04	&	0.487	$\pm$	0.002	&	15.1	$\pm$	0.3	&	10.37	$_{-	0.02	} ^{+	0.02	}$   &	0.51	$_{-	0.07	} ^{+	0.08	}$   &	-					&		\\
53629.38	&	91702-01-79-01	&	0.514	$\pm$	0.002	&	16.6	$\pm$	0.2	&	9.74	$_{-	0.01	} ^{+	0.01	}$   &	0.59	$_{-	0.04	} ^{+	0.04	}$   &	-					&		\\
53630.49	&	91702-01-79-00	&	0.543	$\pm$	0.002	&	18.2	$\pm$	0.1	&	8.68	$_{-	0.02	} ^{+	0.02	}$   &	0.48	$_{-	0.04	} ^{+	0.04	}$   &	-					&		\\
53631.47	&	91702-01-80-00	&	0.584	$\pm$	0.002	&	20.3	$\pm$	0.1	&	7.76	$_{-	0.02	} ^{+	0.02	}$   &	0.55	$_{-	0.05	} ^{+	0.05	}$   &	-					&		\\
53632.45	&	91702-01-80-01	&	0.667	$\pm$	0.002	&	24.0	$\pm$	0.1	&	4.736	$_{-	0.008	} ^{+	0.008	}$   &	0.62	$_{-	0.04	} ^{+	0.04	}$   &	-					&		\\
53633.50	&	91702-01-81-00	&	0.736	$\pm$	0.004	&	25.1	$\pm$	0.1	&	2.15	$_{-	0.02	} ^{+	0.02	}$   &	0.33	$_{-	0.05	} ^{+	0.05	}$   &	-					&		\\
53634.11	&	91702-01-80-02	&	0.749	$\pm$	0.006	&	25.1	$\pm$	0.4	&	1.44	$_{-	0.02	} ^{+	0.02	}$   &	0.29	$_{-	0.05	} ^{+	0.06	}$   &	-					&		\\
53634.31	&	91702-01-81-01	&	0.758	$\pm$	0.004	&	25.7	$\pm$	0.2	&	1.32	$_{-	0.01	} ^{+	0.006	}$   &	0.24	$_{-	0.03	} ^{+	0.02	}$   &	-					&		\\
53635.47	&	91702-01-81-02	&	0.782	$\pm$	0.004	&	25.2	$\pm$	0.2	&	0.584	$_{-	0.003	} ^{+	0.007	}$   &	0.09	$_{-	0.01	} ^{+	0.01	}$   &	9	$_{-	1	} ^{+	1	}$ &	$L_{l}$	\\
53636.19	&	91702-01-87-03	&	0.785	$\pm$	0.005	&	25.4	$\pm$	0.4	&	0.46	$_{-	0.01	} ^{+	0.01	}$   &	0.10	$_{-	0.03	} ^{+	0.04	}$   &	12	$_{-	2	} ^{+	3	}$ &	$L_{l}$	\\
53636.45	&	91702-01-82-00	&	0.867	$\pm$	0.005	&	24.3	$\pm$	0.5	&	0.43	$_{-	0.02	} ^{+	0.01	}$   &	0.13	$_{-	0.05	} ^{+	0.05	}$   &	-					&		\\
53637.17	&	91704-01-01-00	&	0.787	$\pm$	0.006	&	25.4	$\pm$	0.4	&	0.34	$_{-	0.01	} ^{+	0.003	}$   &	0.03	$_{-	0.02	} ^{+	0.07	}$   &	8	$_{-	2	} ^{+	2	}$ &	$L_{l}$	\\
53637.24	&	91704-01-01-01	&	0.792	$\pm$	0.004	&	25.6	$\pm$	0.2	&	0.316	$_{-	0.002	} ^{+	0.002	}$   &	0.070	$_{-	0.008	} ^{+	0.009	}$   &	6.1	$_{-	0.5	} ^{+	1.2	}$ &	$L_{l}$	\\
53637.50	&	91704-01-01-02	&	0.796	$\pm$	0.006	&	24.7	$\pm$	0.4	&	0.31	$_{-	0.01	} ^{+	0.005	}$   &	0.04	$_{-	0.01	} ^{+	0.04	}$   &	4.8	$_{-	0.6	} ^{+	0.8	}$ &	$L_{l}$	\\
53638.35	&	91702-01-86-00	&	0.791	$\pm$	0.005	&	25.8	$\pm$	0.4	&	0.231	$_{-	0.007	} ^{+	0.004	}$   &	0.05	$_{-	0.02	} ^{+	0.02	}$   &	4.3	$_{-	0.6	} ^{+	0.7	}$ &	$L_{l}$	\\
53639.14	&	91702-01-86-01	&	0.805	$\pm$	0.007	&	25.6	$\pm$	0.5	&	0.197	$_{-	0.006	} ^{+	0.02	}$   &	0.06	$_{-	0.02	} ^{+	0.02	}$   &	4.8	$_{-	0.9	} ^{+	0.9	}$ &	$L_{l}$	\\
53639.20	&	91702-01-86-04	&	0.821	$\pm$	0.008	&	26.6	$\pm$	0.7	&	0.2	$_{-	0.0	} ^{+	0.005	}$   &	0.04	$_{-	0.01	} ^{+	0.04	}$   &	5	$_{-	1	} ^{+	2	}$ &	$L_{l}$	\\			
\hline	
\end{longtable}
\end{landscape}
\end{center} 																																														
%%_____________________END________SAMPLEA____________________________%%
		
\begin{center}                                                                                                                                                                                                                                  \begin{landscape}
%\begin{footnotesize}
\begin{longtable}{|c c c  | c c c c | c c c c | c c c c|}                                                                                                                                                                     
\caption{Observations included in Sample B (see Sec. \ref{sec:selection}). For each observation we report the Observation ID, the time at which the observation was taken, the integrated fractional rms (measured in the 2-27 keV energy band and in the 0.0-64.0 Hz frequency range), the hardness ratio calculated as described in Sec. \ref{sec:observations}, the centroid frequency of the Type-C QPO, of the lower and of the upper HFQPO, their width, their fractional rms and their single trial significance. \bigskip}\label{tab:sampleB} \\											

\hline                                                                                                                                                                                                                                                  \hline                                                                                                                                                                                                                                                 

\multicolumn{15}{c}{{\bf Sample B observations}} \\                                                                                                                                   \hline
\hline
\multicolumn{15}{c}{ } \\                                                                                                                                   \hline                                                                                                                                                                                                                                                  \multicolumn{15}{|c|}{Sample B1 observations}\\                                                                                                                                   																										
\hline																						
\hline																			
\multicolumn{3}{|c|}{} & \multicolumn{4}{|c|}{Type-C QPO} & \multicolumn{4}{|c|}{lower HFQPO} & \multicolumn{3}{|c|}{upper HFQPO} \\                                                                                                                                                                                                                                                            																										\hline                                                                                                                                                                                                                       \hline                                                                                                                                                                                                                                                                                                                                                                                                                                                                  			
		Obs ID	&		RMS		&	Hardness			&		$\nu$						&		$\Delta \nu$						&	rms							&	sign.	&		$\nu$						&		$\Delta \nu$						&	rms							&	sign.	&		$\nu$						&		$\Delta \nu$						&	rms							&	sign.	\\
			&			fractional (\%)		&				&		[Hz]						&		[Hz]						&	\%							&	[$\sigma$]	&		[Hz]						&		[Hz]						&	\%							&	[$\sigma$]	&		[Hz]						&		[Hz]						&	\%							&	[$\sigma$]	\\
\hline																												\hline																														10255-01-06-01	&		5.08	$\pm$	0.03	&	0.472	$\pm$	0.002	&								&								&								&		&								&								&								&		&								&								&								&		\\
		10255-01-07-00	&		4.37	$\pm$	0.02	&	0.442	$\pm$	0.001	&	$	17.3	_{	-0.1	}^{	0.1	}$	&	$	3.9	_{	-0.2	}^{	0.2	}$	&	$	1.62	_{	-0.06	}^{	0.05	}$	&	13	&	$	298	_{	-4	}^{	4	}$	&	$	35	_{	-14	}^{	22	}$	&	$	0.59	_{	-0.05	}^{	0.05	}$	&	6	&	$	441	_{	-2	}^{	2	}$	&	$	30	_{	-6	}^{	8	}$	&	$	4.5	_{	-0.3	}^{	0.3	}$	&	7	\\
		10255-01-17-00	&		4.36	$\pm$	0.02	&	0.461	$\pm$	0.001	&								&								&								&		&								&								&								&		&								&								&								&		\\
\hline																																																			\multicolumn{15}{c}{ } \\                                                                                                                                  \hline                                                                                                                                                                                                                                                  																																																							
\multicolumn{15}{|c|}{Sample B2 observation} \\                                                                                                                                   																															
\hline																												\hline
																														Obs ID	&			RMS		&	Hardness			&		$\nu$						&		$\Delta \nu$						&	rms							&	sign.	&		$\nu$						&		$\Delta \nu$						&	rms							&	sign.	&		$\nu$						&		$\Delta \nu$						&	rms							&	sign.	\\
			&			fractional (\%)		&				&		[Hz]						&		[Hz]						&	\%							&	[$\sigma$]	&		[Hz]						&		[Hz]						&	\%							&	[$\sigma$]	&		[Hz]						&		[Hz]						&	\%							&	[$\sigma$]	\\
\hline																												\hline																																	
		10255-01-09-00	&		4.67	$\pm$	0.02	&	0.418	$\pm$	0.001	&	$	18.3	_{	-0.1	}^{	0.1	}$	&	$	5.2	_{	-0.3	}^{	0.3	}$	&	$	1.30	_{	-0.03	}^{	0.06	}$	&	22	&	-							&	-							&	-							&	-	&	$	451	_{	-5	}^{	6	}$	&	$	36	_{	-8	}^{	12	}$	&	$	4.6	_{	-0.4	}^{	0.4	}$	&	5	\\
		10255-01-10-00	&		4.35	$\pm$	0.02	&	0.428	$\pm$	0.001	&	$	18.1	_{	-0.1	}^{	0.1	}$	&	$	5.2	_{	-0.4	}^{	0.4	}$	&	$	1.44	_{	-0.06	}^{	0.06	}$	&	12	&	-							&	-							&	-							&	-	&	$	446	_{	-4	}^{	4	}$	&	$	36	_{	-7	}^{	9	}$	&	$	5.3	_{	-0.4	}^{	0.5	}$	&	6	\\
\hline

\end{longtable}
%\end{footnotesize}
\end{landscape}																					\end{center}

%\end{landscape}
%%_____________________END__________SAMPLEB____________________________%%
\normalsize

%%_____________________BEGIN________DISTPAR____________________________%%

\begin{table*}
\begin{center}                                                                                                                                                                      \caption{Black hole parameters and associated quantities as measured from the RPM. For details, see the text, Sec. \ref{sec:RPM_solve}.\bigskip}\label{tab:distribution_par}						
\begin{tabular}{|c c c|}	

\hline                                                                                                                                                                                                                                            

 	&	Mean value	&	Standard deviation	\\	
\hline                                                                                                                                                                                                                                                  						
\hline                                                                                                                                                                                                                                                  						
Mass (Solar masses)	&	5.307	&	0.066	\\	
Spin	&	0.286	&	0.003	\\	
Radius (Gravitational radii)	&	5.677	&	0.035	\\	
\hline						
r$_{\rm ISCO}$ (Gravitational radii)	&	5.031	&	0.009	\\	
$\nu_{nod}$ at r$_{\rm ISCO}$ (Hz)	&	24.680	&	0.558	\\	
\hline                                                                                                                                                                                                                                                	
\end{tabular}						
\end{center} 						
\end{table*} 						
 
%%_____________________END__________DISTPAR____________________________%%

%%_____________________START__________WIDTH____________________________%%
\begin{table}
\begin{center}                                                                                                                                                                      \caption{Comparison between the simulated QPO widths and the observed QPO widths (see Sec. \ref{sec:width} for details on the simulation). The observed width corresponds to the values given in Tab. \ref{tab:sampleB}, sub-sample B1. However, for sake of clarity, we give here the 1-sigma confidence interval on the widths measurement. For the case of Type-C QPO we report in this table the minimum observed width and the maximum observed width (the 1-sigma errors are taken into account) of the individual Type-C QPOs in order to allow the comparison avoiding the biases described in the text.\bigskip}\label{tab:confronto_width}		
			
\begin{tabular}{|c c c|}		
			
\hline                                                                                                                                                                                                                                                 

QPO type 	&	simulated width ($\Delta \nu$)	&	Observed width	\\	
			& 	$[Hz]$							&	$[Hz]$			\\	
\hline                                                                                                                                                                                                                                              \hline                                                                                                                                                                                                                                                  						
Type-C QPO	&	2.11 - 2.90		&	2.1 - 4.2			\\	
Upper HFQPO &   41.58 - 57.66	&	21.54 - 57.70		\\	
Lower HFQPO &	26.77 - 36.83	&	24.06 - 37.74		\\	
\hline						
		
\end{tabular}						
\end{center} 						
\end{table}

%%_____________________END__________WIDTH____________________________%%

\label{lastpage}
\end{document}